\documentclass[showpacs,aps,prb,preprint,superscriptaddress]{revtex4}

\usepackage{graphicx}
\usepackage{subfigure}
\usepackage{amsfonts}
\usepackage{float}
\begin{document} 

\title{Pinning and elastic properties of vortex matter in highly strained Nb$_{75}$Zr$_{25}$: Analogy with viscous flow of disordered solids} 

\author{Jagdish Chandra}
\affiliation{Magnetic and Superconducting Materials Section}
\email{megh@rrcat.gov.in}
\author{Meghmalhar Manekar$^*$}
\author{V. K. Sharma}
\affiliation{Magnetic and Superconducting Materials Section}
\author{Puspen Mondal}
\author{Pragya Tiwari}
\affiliation{Indus Synchrotrons Utilisation Division\\ Raja Ramanna Centre for Advanced Technology, Indore 452 013, India}
\author{S. B. Roy}
\affiliation{Magnetic and Superconducting Materials Section}

\date{\today}
\begin{abstract}

We present the results of magnetization and magneto-transport measurements in the superconducting state of an as-cast Nb$_{75}$Zr$_{25}$ alloy. We also report the careful investigation of the microstructure of our sample at various length scales by using optical, scanning electron and transmission electron microscopies. The information of microstructure is used to understand the flux pinning properties in the superconducting state within the framework of collective pinning. The magneto-transport measurements show a non-Arrhenius behaviour of the temperature and field dependent resistivity in the flux flow region. This non-Arrhenius behaviour is understood in terms of a model, which was originally proposed for viscous flow of disordered solids and is popularly known in the literature as the `shoving' model. The activation energy for flux flow is obtained from magneto-transport measurements and is assumed to be mainly the elastic energy stored in the flux-line lattice. The critical current density estimated from magnetization measurements is moderately high of the order of 10$^8$ Am$^{-2}$ at a temperature of 2K. The scaling of pinning force density with respect to reduced field indicates the presence of two pinning mechanisms of different origins.  The elastic constants of the flux-line lattice are estimated from magnetization measurements and are used to estimate the length scale of vortex lattice movement, or the volume displaced by the flux-line lattice, during flux flow. It appears that the vortex lattice displacement estimated from elastic energy considerations is of the same order of magnitude as that of the flux-bundle hopping length when a finite resistance appears during flux flow. Our results could provide possible directions for establishing a framework where vortex matter and glass forming liquids or amorphous solids can be treated in a similar manner for understanding the phenomenon of viscous flow in disordered solids or more generally the pinning and depinning properties of elastic manifolds in random media.

\end{abstract}

\maketitle 

\section{Introduction}

The phenomena of flux creep\cite{creep} and thermally activated flux flow\cite{flow} were initially discovered in conventional hard type-II superconductors. The ``irreversibility line" in the $H-T$ plane, which separates the reversible (or flux flow) and irreversible magnetization (or flux pinned) regions, was thought to be exclusive to the high-temperature superconductors and was thus interpreted as a ``quasi de Almeida-Thouless line" in analogy to the irreversibility observed between the zero-field-cooled (ZFC) and field-cooled (FC) magnetization in spin glasses and the possibility of a superconducting glass phase was conjectured.\cite{muller} The vortex-glass phase was suggested to be a new thermodynamic phase within the mixed state of a type-II superconductor.\cite{fisher} The term vortex glass was used to the signify the destruction of long range translational order in the Abrikosov lattice due to flux pinning by the underlying disorder in the superconducting material. The irreversibility line was shown to be a sharp boundary across a thermodynamic phase transition between the disordered vortex solid (or glass) and the vortex liquid where thermally activated flux flow is possible.\cite{fisher} An alternate interpretation of the irreversibility line, without invoking the glass phase in the mixed state, was later offered in terms of the more conventional phenomenon like the flux creep.\cite{yeshurun} These paradigms of high-$T_C$ superconductors were applied to conventional type-II materials like niobium thin films where it was found that the irreversibility line represents the vortex-melting transition.\cite{paradigm} The model of the ``giant flux creep"\cite{yeshurun} was extended to explain the \emph{width} and \emph{shape} of the resistive transition as a function of applied magnetic field.\cite{tinkham} The key ingredient to the flux creep picture is the estimate of the activation energy $U_0$ which must be surmounted for the flux movement to occur.\cite{yeshurun} The elemental volume of the flux line lattice which moves during the creep is governed by the activation energy and thus gives the temperature width of the resistive transition as a function of applied field.\cite{tinkham} The superconducting glass picture\cite{muller} and the giant flux creep picture\cite{yeshurun} were thus shown to be \emph{qualitatively} similar by accounting for the field dependence of width of transition (and in-turn the shape of irreversibility line) without the need for invoking sample inhomogeneities.\cite{tinkham} The combination of quenched disorder, higher temperatures, shorter coherence length and large magnetic penetration depth in case of high T$_C$ superconductors lead to a modification of the mean field phase diagram of the conventional low T$_C$ superconductors.\cite{ffh} 

An alternate way of studying the vortex matter is to consider the collection of vortices as an elastic object,\cite{vortex-phases} instead of treating the problem within the framework of microscopic Ginzburg-Landau phenomenology.\cite{ffh} The rich variety of vortex phases within the superconducting state could be well explained by an elastic theory of vortex matter.\cite{giamarchi-prl,giamarchi-prb} By supplementing the traditional theory of collective pinning\cite{larkin} with the concepts of elastic media in random potential, the various properties of the vortex matter could be described in an unified manner.\cite{blatter-rmp} The application of the theory of elasticity to vortex matter has found similarities in case of other systems like magnetic domain walls, charge density waves and wigner crystals.\cite{vortex-phases} The vortex matter has thus provided a model system to study the properties of moving glasses or driven elastic media in presence of quenched disorder.\cite{gia-moving-prl,moving-prb}  

In the work presented here, we experimentally attempt to find similarities between the flux flow in vortex matter with one more class of a driven elastic system, namely the flow of a molecular glass or viscous liquid. We try to address one of the fundamental questions arising in the studies of vortex matter, which is, ``How much distance does the fluxon (or the elemental volume of flux-line lattice) move for a finite resistance to appear across the resistive transition in a hard type-II superconductor?" While this question can be of technological importance for tuning the nature of quenched disorder to achieve desirable critical current densities, it can also lead to fundamental understanding of elastic systems influenced by disorder and driven by external force. We have chosen the Nb$_{75}$Zr$_{25}$ alloy for this study. The Nb-Zr alloy system had generated considerable interest in the past for its perceived use in superconducting wires for high field magnets.\cite{whatever} The $\beta$ phase alloy with the b.c.c. structure was well studied due to its high critical current carrying ability.\cite{whatever} The same composition in this alloy system was used to investigate fundamental phenomenon like the flux creep in hard superconductors\cite{creep} or the scaling laws for flux pinning in hard type-II superconductors,\cite{kramer} which are now known popularly as Kramer scaling.    

Our approach to solve the above posed question is as follows. We first characterize the microsctructure at various length scales of the Nb$_{75}$Zr$_{25}$ superconducting alloy under consideration to know about the nature of disorder which can act as pinning potential for the flux-line lattice. The estimate of the activation energy required for flux-line movement is obtained from electrical resistivity measurements across the transition as a function of both temperature and magnetic field. The non-Arrhenius shape of the resistive transition draws our attention to a model of viscous flow which has been used earlier to explain a non-Arrhenius temperature dependent viscosity across a molecular glass transition. The model is based on purely elastic energy considerations where the activation energy is related to the shear modulus through some sort of a \emph{correlation} volume and is popularly known as the \emph{shoving} model.\cite{shoving} The field-dependent pinning force in the mixed state of our sample can be explained only if more than one pinning mechanisms are considered. Such a possibility is explored within the framework of collective pinning,\cite{larkin} where the estimates of correlation lengths over which the flux bundles are pinned can be obtained. The elastic constants of vortex matter like the shear modulus ($C_{66}$) and the tilt modulus ($C_{44}$) are obtained from magnetization measurements using this framework. The activation energy for flux-line movement and the elastic constants of the vortex matter are then correlated through the correlation volume of flux bundle to get the estimate of the distance over which the flux-line lattice moves before a finite resistance appears across the transition. Interestingly, we find that the displacement of vortex matter estimated from the shoving model is of the same order of magnitude as that of the flux-bundle hopping length when a finite resistance appears across the resistive transition. The viscous flow of disordered solids and the flux-line movement in vortex matter appear to be just different cases of the same general phenomenon. These results could possibly lead to newer directions for encompassing diverse phenomenon arising in periodic systems influenced by quenched disorder with vortex matter as a model system.

\section{Experimental}

Nb$_{75}$Zr$_{25}$ alloy ($\sim$ 1gm mass) was prepared by arc melting the constituent elements of 99.99\% purity in a water cooled copper hearth placed in an inert argon atmosphere. The resulting sample button was remelted six times to ensure homogeneity and was not subjected to any further heat treatment. The sample was then characterized by x-ray diffraction (XRD) using a commercial diffractometer (Panalytical X-Pert PRO MRD) with Cu-K$_\alpha$ radiation. $\theta$-2$\theta$ scans were recorded in the Bragg-Brentano geometry. A monochromatized (Cu-K$_{\alpha1}$) and collimated (about 20 arc-sec in the plane of scattering) x-ray beam was obtained using a hybrid monochromator. The obtained x-ray peak widths, which were significantly larger than the instrumental broadening, were used for the estimation of particle size and lattice strain. A small piece of sample was used for optical metallography which was subjected to slow grinding using SiC paper and polishing using diamond compound on a soft cloth. After obtaining a mirror quality finish on the surface, the sample was chemically etched using a mixture of 70 ml C$_3$H$_6$O$_3$ (lactic acid) with 30 ml HNO$_3$ (nitric acid) and 2 ml HF (hydrofluoric acid) for about 15 seconds. The microstructure was observed with a commercial inverted metallurgical optical microscope (Leica DMI5000M). A commercial scanning electron microscope (Philips, XL30CP) equipped with a energy dispersive spectrometer (Bruker, XFlash$\circledR$  Silicon Drift Detector) was used for observing the microstructure on sub-micron length scales and determining the chemical composition at various locations of the sample.  For visualizing the disorder/grain structure at even smaller length scales, a small piece of the sample was cut from the same parent button for transmission electron microscopy (TEM) studies. A commercial TEM (Phillips, CM200) with a tungsten filament as cathode, was used at an accelerating anode voltage of 200 kV. Electrical resistivity was measured using the standard four-probe technique with a home-made variable temperature insert in a commercial superconducting magnet and cryostat system (American Magnetics Inc., USA) Magnetization ($M$) measurements were performed as a function of field ($H$) and temperature ($T$) using a commercial vibrating sample magnetometer (VSM Quantum Design, USA). We consistently use the SI units for all the physical properties reported here. 

\section{Results and discussion}

\subsection{Microstructure and nature of disorder}

\begin{figure*}
\centering
\subfigure[]{
\includegraphics[width=7cm,height=6cm]{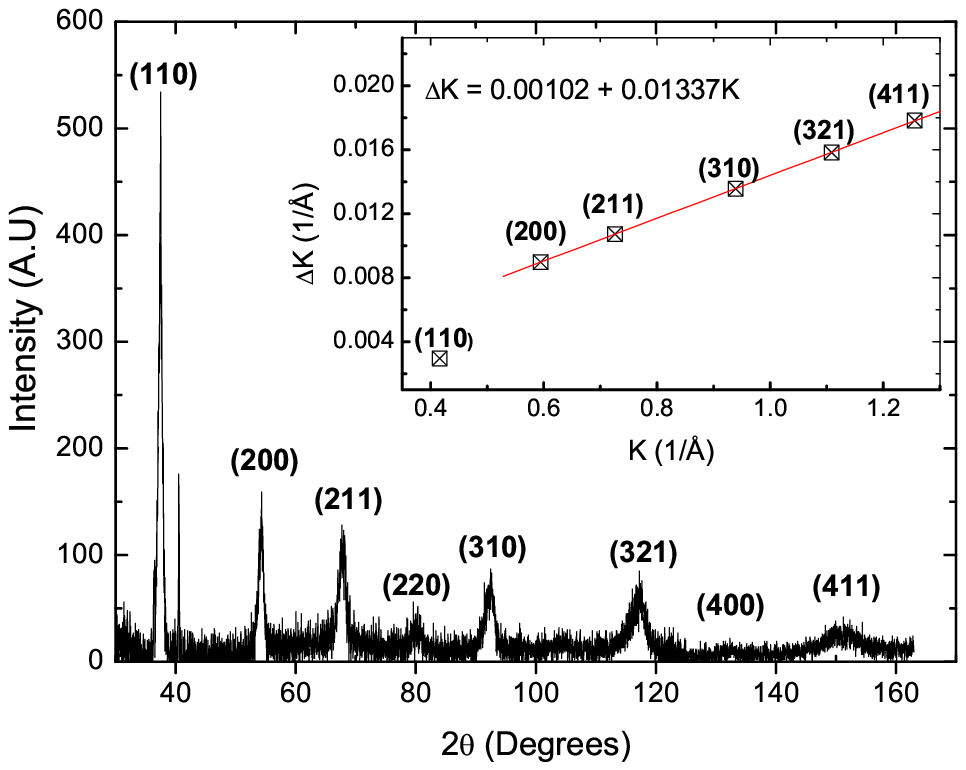}
\label{xrd}}
\hspace{0.5cm}
\subfigure[]{
\includegraphics[width=7cm,height=6cm]{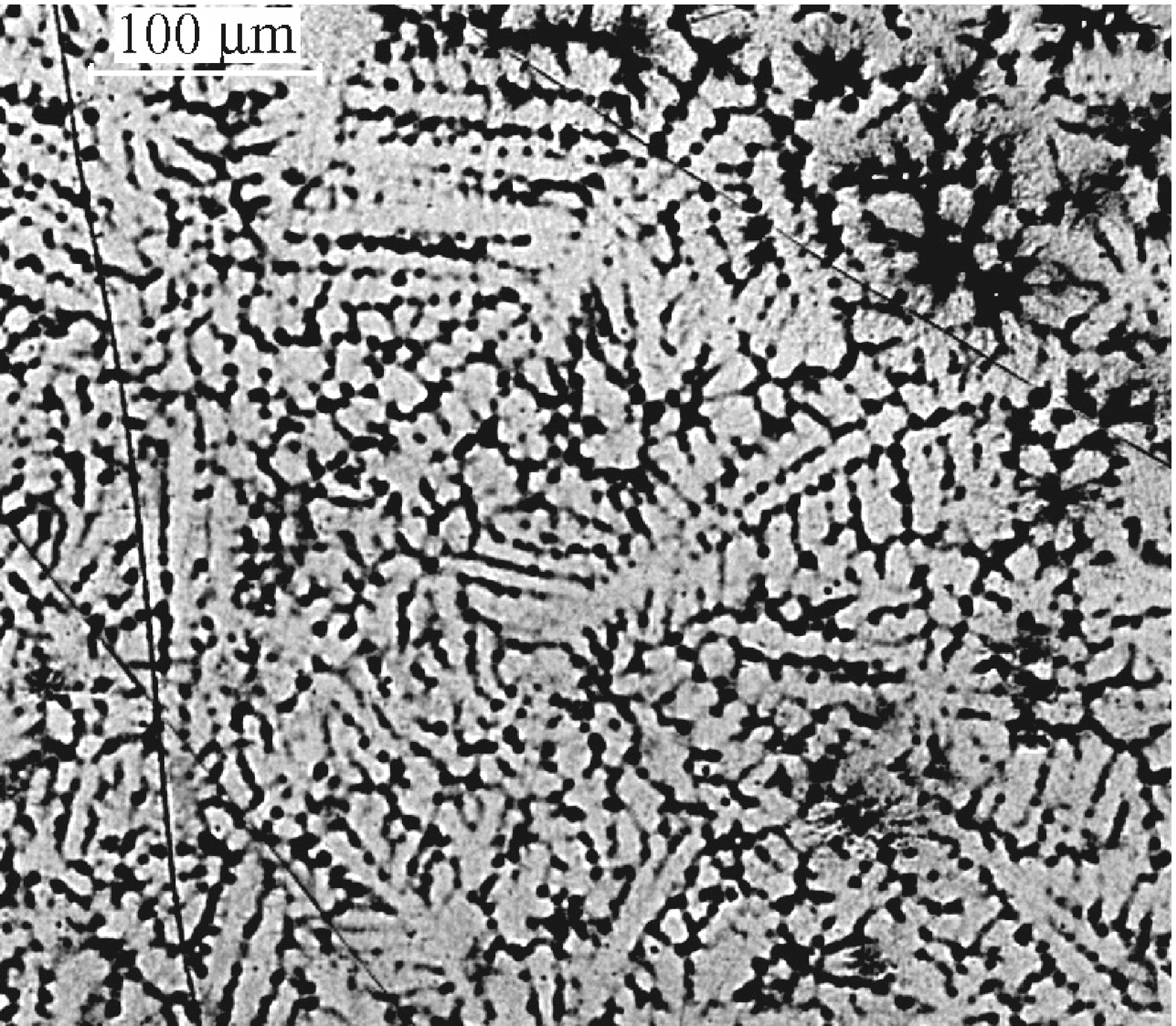}
\label{micro}}
\caption{\label{xrd-micro}(a) The x-ray diffraction pattern of Nb$_{75}$Zr$_{25}$ as-cast alloy. Inset shows the Williamson-Hall plot. (b) The microstructure of the same sample observed under an optical microscope. The dendritic pattern is clearly visible.}
\end{figure*}

Figure \ref{xrd} shows the XRD pattern of the sample. The peaks which are quite broad in nature could be indexed with the $\beta$ phase b.c.c. structure with a lattice constant of 3.368 \AA. To estimate the grain size and lattice strain, we use the Williamson-Hall (W-H) plot,\cite{wh-plot} which is shown in the inset of fig. \ref{xrd}.  The plot is between $\Delta K=(cos(\theta) \Delta(\theta))/\lambda$ and $K=sin(\theta)/\lambda$, where $\Delta(\theta)$ is the full width at half maximum for each of the Bragg peaks. The (110) peak appears to be anomalously sharp and does not belong to the straight line W-H fit obtained from the other peaks. The reciprocal of the y-intersept gives the average grain size which is estimated to be around 98 nm. The slope of the W-H plot gives the lattice strain which is found to be about 1.33\%. The microstructure (see fig. \ref{micro}) shows a dendritic growth during the solidification of the alloy from melt, which is quite commonly seen in many materials when a thermal gradient is present between the melt and the solidified portion of the material.\cite{cahn} The combined results of XRD and optical metallography show that the alloy has probably solidified with substantial disorder due to the rapid cooling it had to experience because of the small size of the sample button placed in the water cooled copper hearth. The rapid cooling probably did not allow sufficient time for the phase to grow homogeneously. 

\begin{figure*}
\centering
\subfigure[]{
\includegraphics[width=7cm]{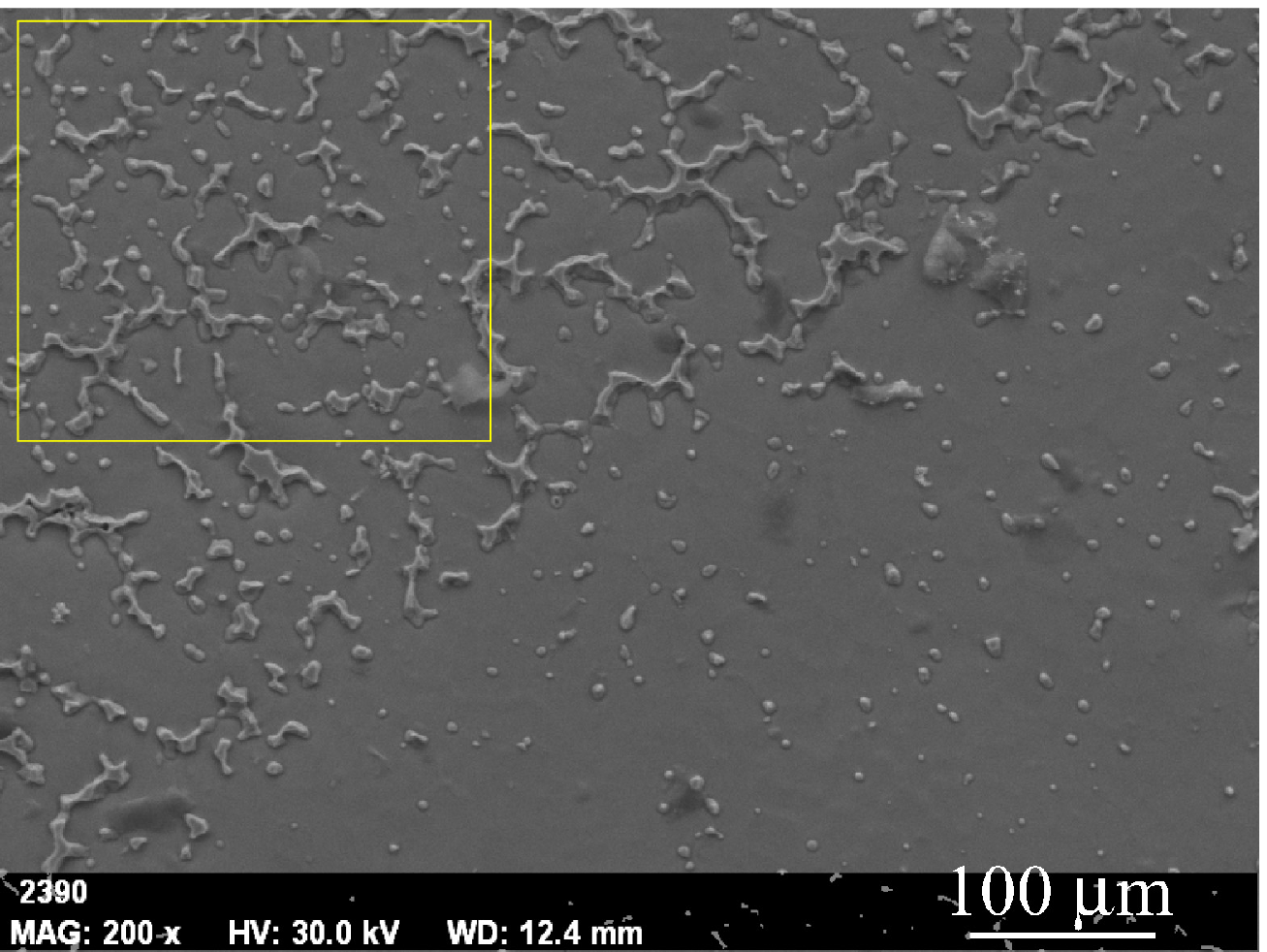}
\label{comp}}
\hspace{0.5cm}
\subfigure[]{
\includegraphics[width=7cm]{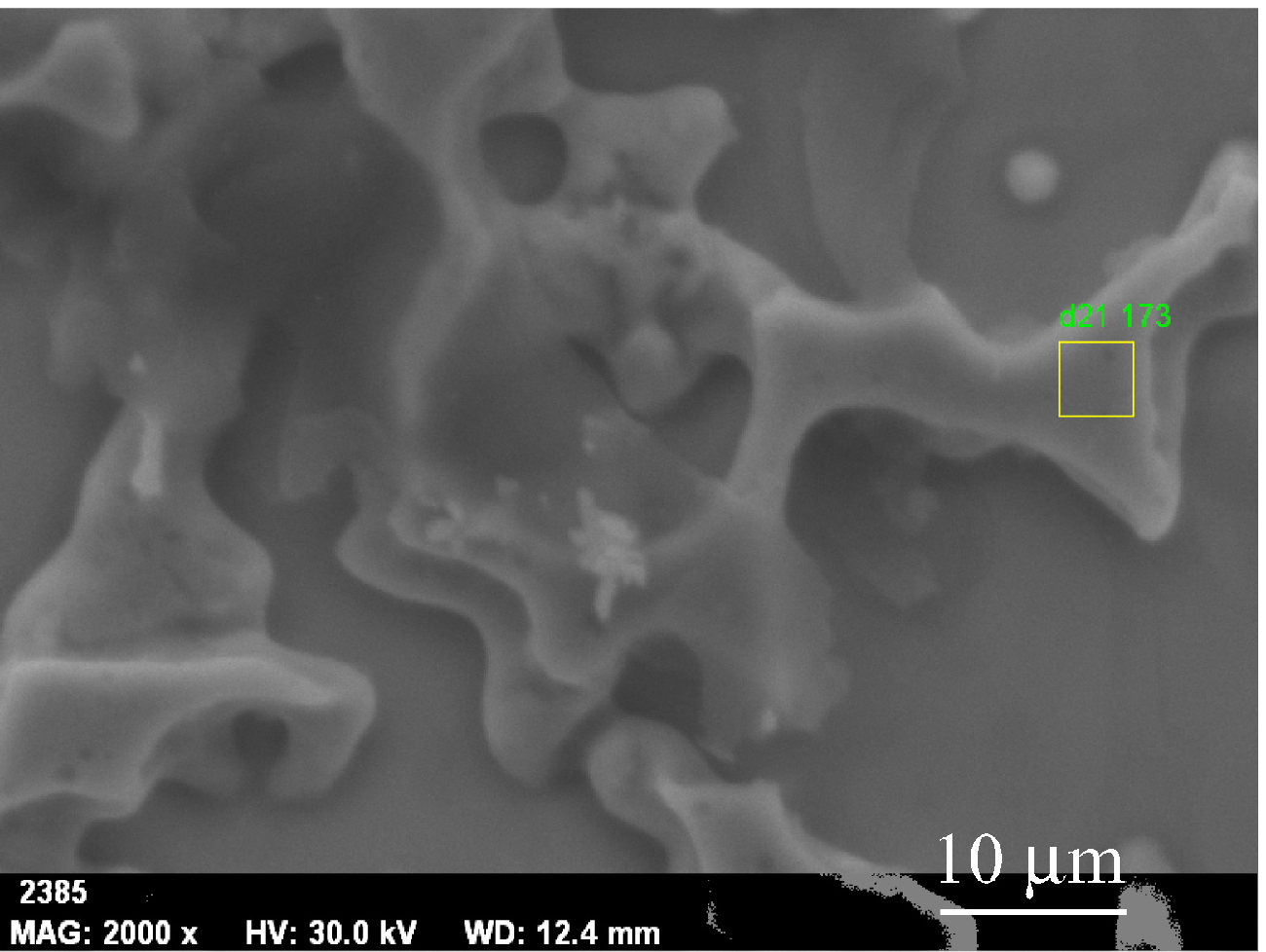}
\label{zr-rich}}
\caption{\label{sem-eds}Scanning electron microscopy images at two different magnifications. (a) The magnification is 200x and the average composition of Nb$_{75}$Zr$_{25}$is observed over the large area marked as a rectangle. (b) The magnification is 2000x and a region on a dendritic arm is in focus. The area is Zr-rich in composition.}
\end{figure*}

The composition variation across this dendritic pattern was determined using energy dispersive spectroscopy. Figure \ref{sem-eds} shows the microstructure at two different length scales at a random location on the sample surface. When the composition is determined over a larger area (fig. \ref{comp}), it is quite close to the nominal composition of Nb$_{75}$Zr$_{25}$ within the error bar of EDS measurements, which is nearly 3\% for Nb and 1\% for Zr. At certain locations on the protrusions of the dendritic arms (fig. \ref{zr-rich}) the Zr concentration is much higher than the target composition (25\%) and has reached up to 40 \% at certain locations of the sample. These results imply that the sample is a mixture of Zr-rich and Zr-deficient regions on the local scale even though the 3:1 ratio of Nb to Zr is maintained on a larger length scale.   

\begin{figure*}
\centering
\subfigure[]{
\includegraphics[width=5cm]{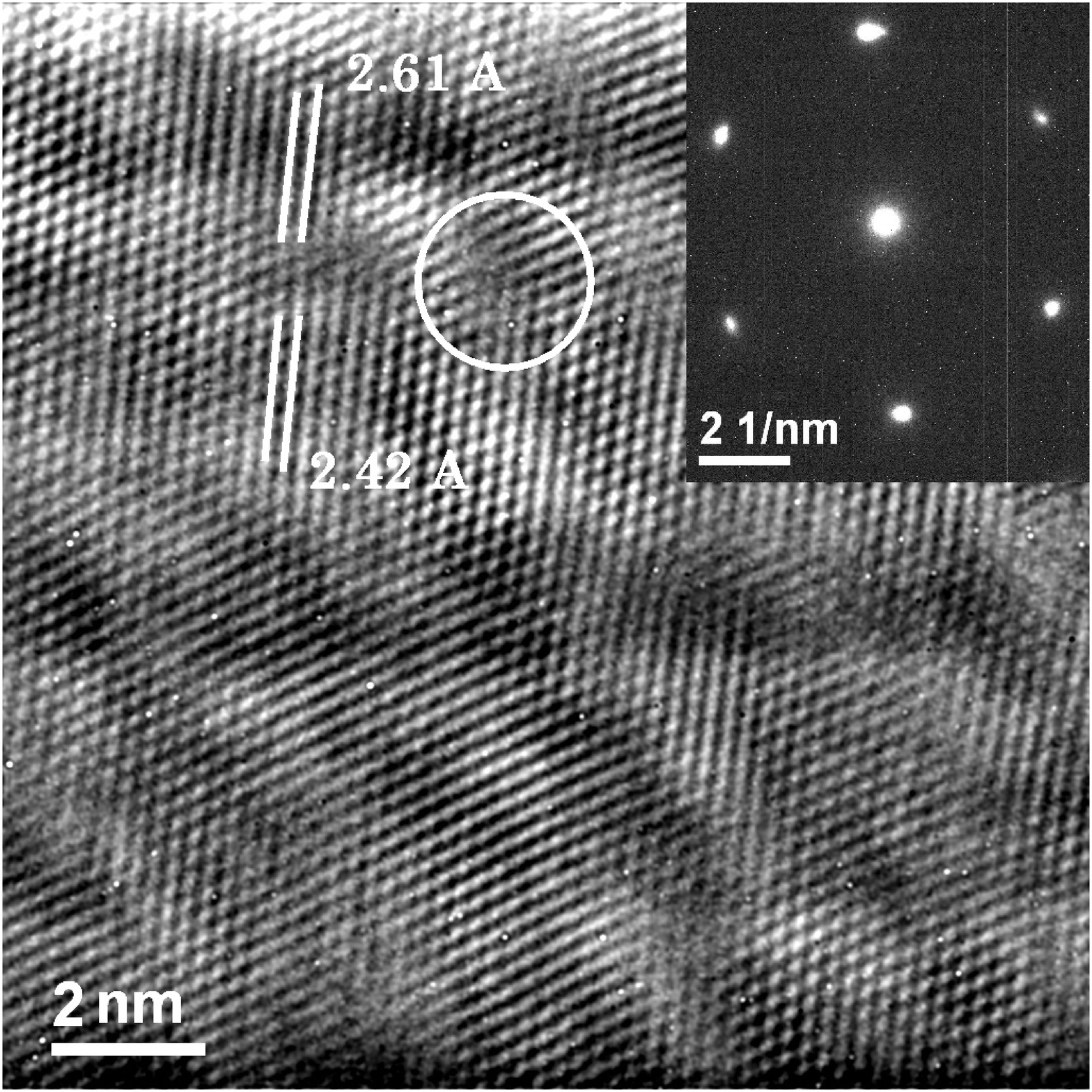}
\label{mismatch}}
\subfigure[]{
\includegraphics[width=5cm]{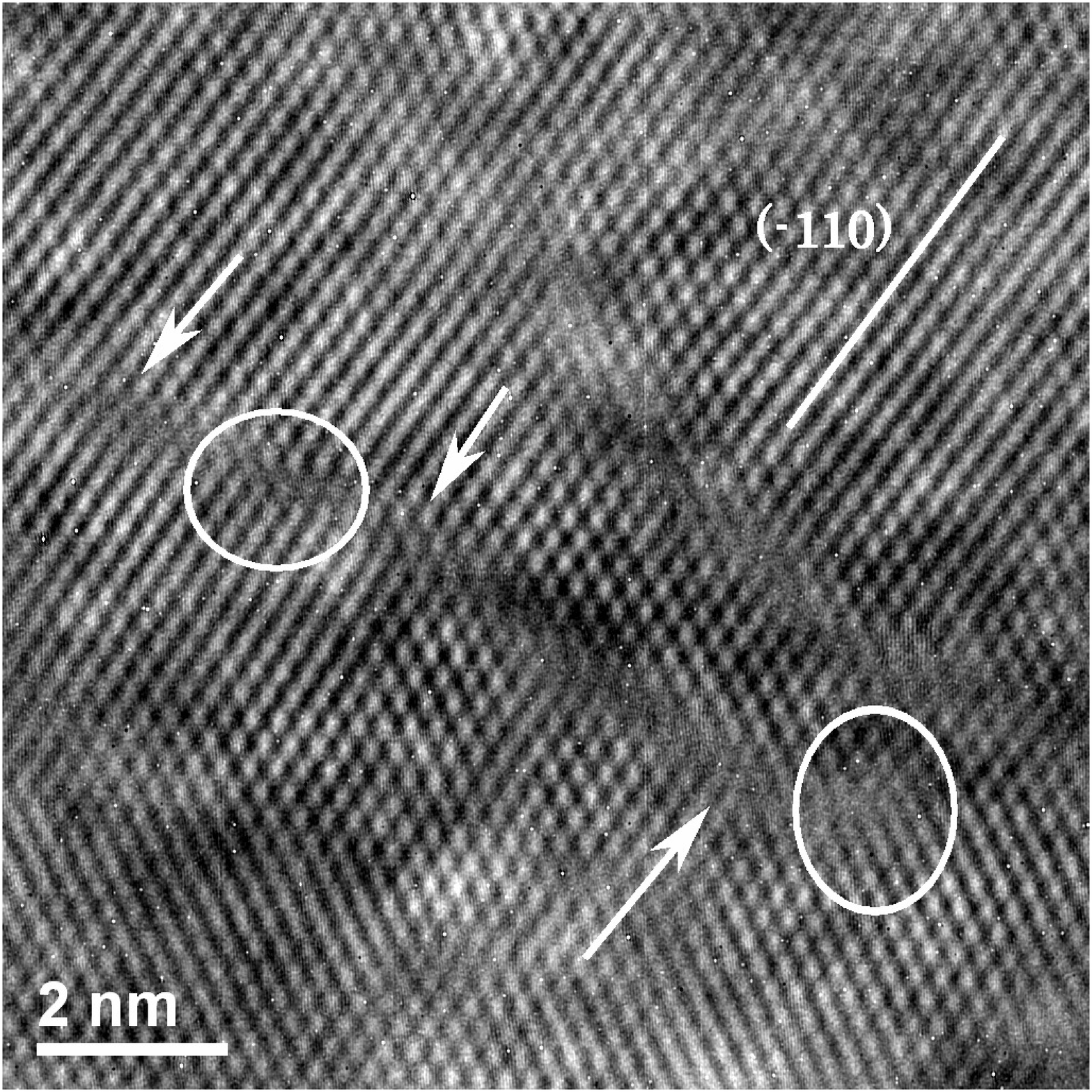}
\label{dislocation}}
\subfigure[]{
\includegraphics[width=5cm]{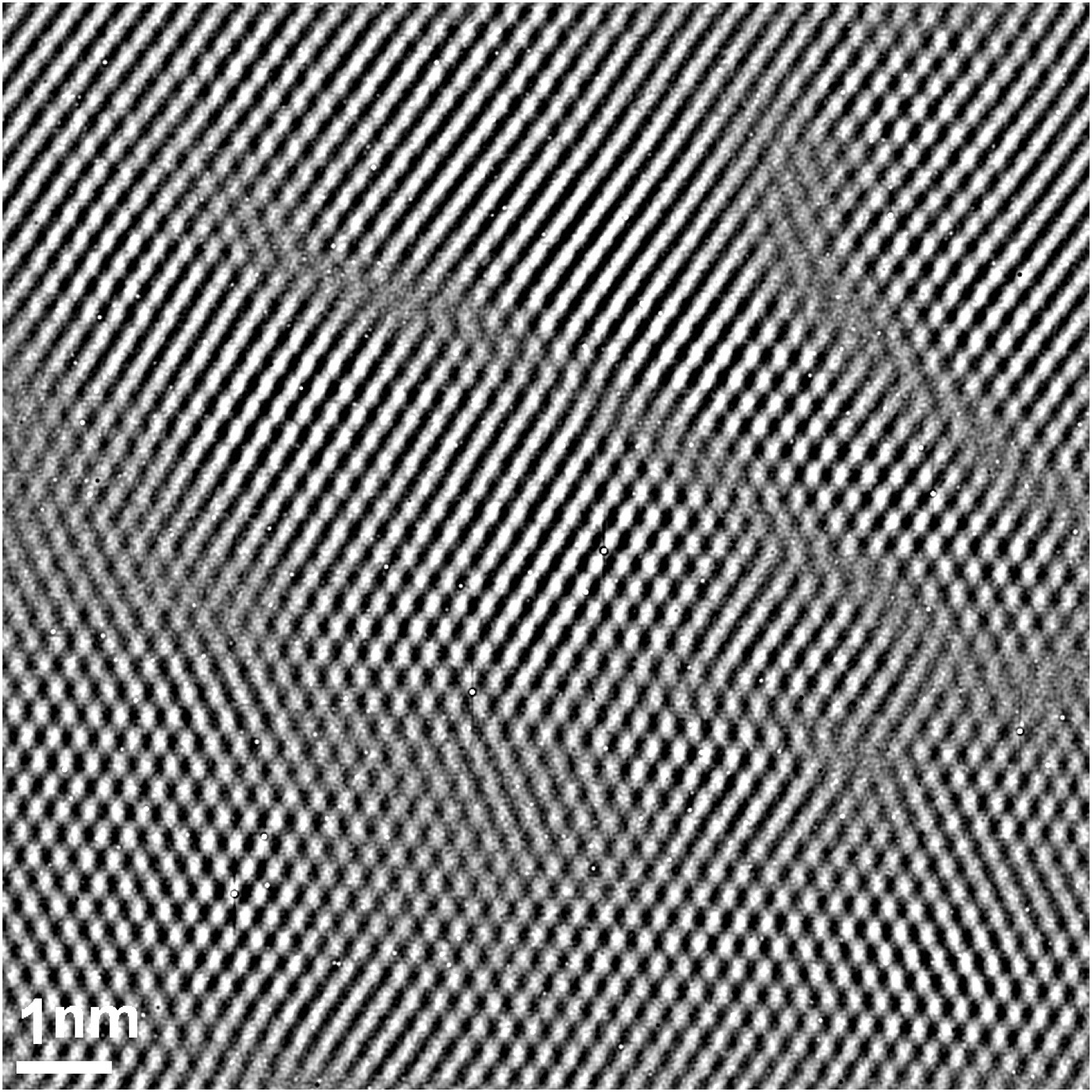}
\label{bending}}
\caption{\label{tem}High resolution transmission electron microscopy images at different locations of the sample. (a) Location where the lattice mismatch and dislocations can be seen. (b) Arrows mark the plane bending and possible jog (middle arrow). Circles mark the edge dislocations. (c) The protrusions of the dendritic growth are amorphous regions that lead to plane bending.}
\end{figure*}

Figure \ref{tem} shows the results of the high resolution TEM studies which bring out the nature of disorder in a more clear manner. In fig. \ref{mismatch} the major defects are lattice plane bending, mismatch in interplanar spacing and edge dislocations. The inset shows the selected area diffraction (SAD) pattern which shows the elongation of Laue spots. It is known that the lattice constant in the Nb-Zr alloy system varies with Zr concentration.\cite{variation} Thus the mismatch in interplanar spacing from 2.61 \AA \ to 2.42 \AA \ at a particular location shown in fig. \ref{mismatch} is probably due to composition variation across that region. This observation is consistent with the results of EDS measurements mentioned earlier. The formation of dislocations (shown in encircled region) is most probably due to a large lattice mismatch at certain locations due to compositional variation. When the strain increases beyond a certain limit in heterostructures, the interface accommodates a part of the lattice mismatch through the introduction of dislocations.\cite{dislocation} Figure \ref{dislocation} shows another location of the sample which contains lattice plane bending and possibly a jog at one of the edge dislocation. Figure \ref{bending} shows a location of the sample within the dendritic growth where the lattice is heavily deformed due to the composition variation. It should be noted that no well defined grains (or grain boundaries) were observed on length scales estimated from x-ray diffraction. The sample is mostly a disordered alloy with intermittent disturbances in the periodicity of lattice at very short length scales without any sharp grain boundaries. As we see later, the typical length scale of these disturbances in the periodicity of lattice is smaller than the coherence length, which makes the collective pinning of the flux line lattice\cite{larkin} more probable than single particle pinning. The applicability of the collective pinning theory will also be shown during the course of this article by estimating certain superconducting parameters.

\subsection{Non-Arrhenius resistive transition and activation energy}

\begin{figure*}
\centering
\subfigure[]{
\includegraphics[width=7cm]{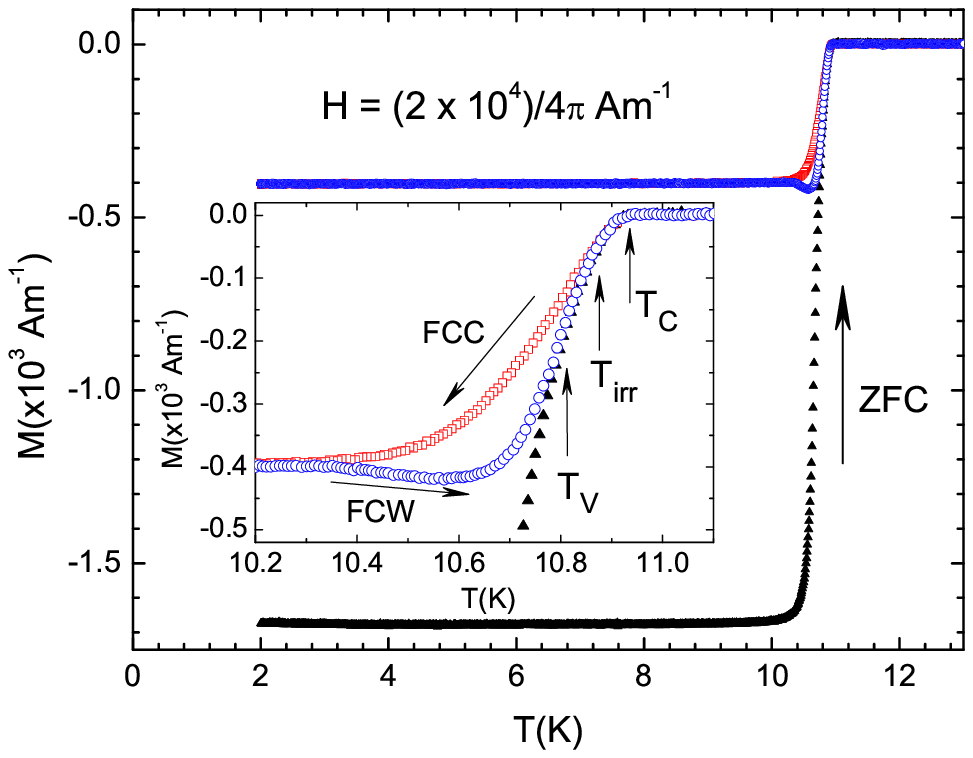}
\label{mt}}
\hspace{0.5cm}
\subfigure[]{
\includegraphics[width=7cm]{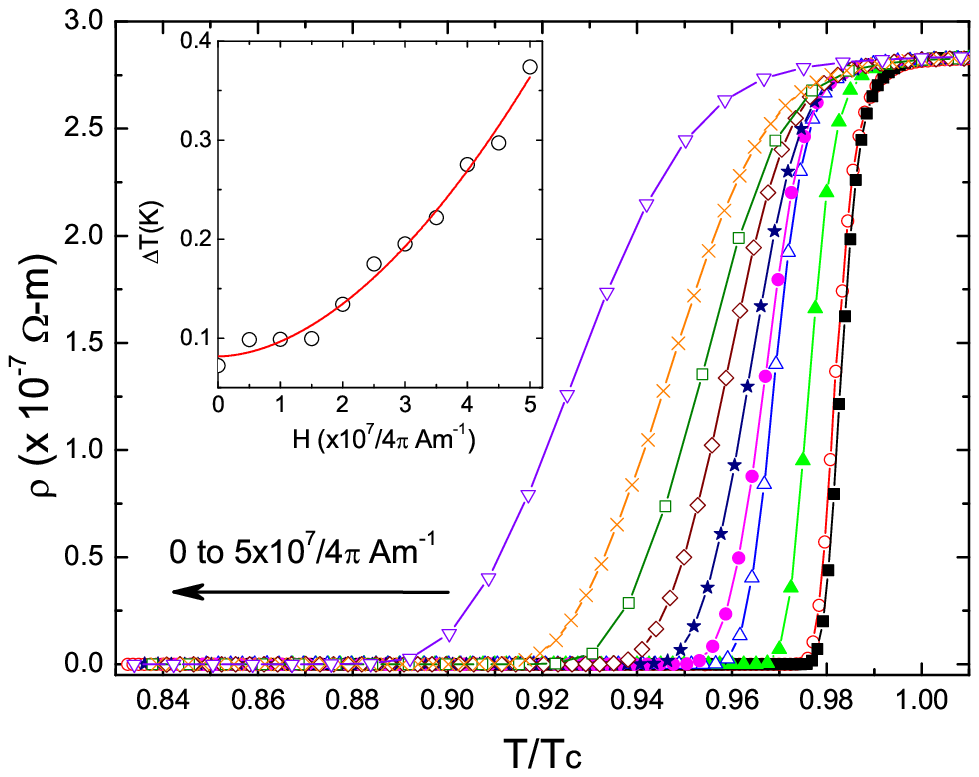}
\label{width}}
\caption{\label{mt-rt}(a) Low field magnetization as a function of temperature in three different thermomagnetic histories. (b) Electrical resistivity as a function of normalized temperature in presence of magnetic field. The lines are a guide to the eye. The width of the transition as a function of magnetic field is shown as the inset. The solid curve in the inset is a fit of experimental data with equation \ref{width-tc}. }
\end{figure*}

The microstructure of the sample at various length scales shows that the material is highly disordered and this inhomogeneity is expected to manifest in the superconducting properties as well. It is known that the $T_C$ in Nb-Zr system varies quite substantially with Zr concentration.\cite{variation} Moreover the lack of long range ordering in the underlying lattice could also result in a response of the sample which is quite different from the bulk of same alloy system. Contrary to this expectation in the case of the present sample, we see that the material behaves like a typical bulk type-II hard superconductor. We highlight this observation before discussing the results of resistivity measurements. Figure \ref{mt} shows the temperature dependent magnetization of the Nb$_{75}$Zr$_{25}$ sample in a low applied magnetic field of $\mu_0 H$ = 2mT. The measurement was performed under three different thermomagnetic histories.

In the zero-field-cooled (ZFC) protocol, the sample is first cooled in zero field down to the lowest temperature. The magnetic field is applied at the lowest temperature and magnetization is measured during warming of the sample. The superconducting transition temperature T$_c$ is estimated to be slightly below 11K. In the field-cooled-cooling (FCC) protocol, the sample is cooled in presence of field from above $T_C$. The temperature at which the FCC curve deviates from the ZFC curve is termed as the irreversibility temperature T$_{irr}$. After reaching the lowest temperature in presence of field, the measurements are performed while warming the sample in field and this protocol is named as the field-cooled-warming (FCW) protocol. The FCW curve deviates from the FCC curve and meets the ZFC curve at a temperature T$_v$ which is slightly lower than T$_{irr}$. These thermal history effects, especially the hysteresis between the FCC and FCW curves, are indications that the sample behaves like a typical bulk hard type-II superconductor.\cite{clem,hyun} We had shown earlier that in case of nanocrystalline superconductors,\cite{nb3al} the FCC and FCW curves may coincide because the grain size is much smaller than the flux trapping depth\cite{clem} inside the sample.  

Figure \ref{width} shows the resistive transition from the normal state to the superconducting state in the of the Nb$_{75}$Zr$_{25}$ sample in various applied fields. The superconducting transition width is less than 0.1K in zero field which is quite unusual for the sample with such composition variation shown earlier. This further confirms that though there is substantial amount of disorder in the sample, it behaves like a bulk superconducting sample of single composition. The width of the transition as a function of field is shown in the inset to fig. \ref{width}. The observed width can be described by the equation,   
\begin{equation}
\Delta T_c = \Delta T_c(0) + H^c
\label{width-tc}
\end{equation}
with $c$ = 1.82. The value of the exponent is not 2/3 as calculated theoretically for a giant flux creep model or superconducting glass.\cite{tinkham} However, it should be noted that the quantitative dependence of the transition width on the applied field depends on the nature of activation energy.\cite{tinkham} The quantitative predictions of the theory may actually depend on the fluxon jumping length.\cite{tinkham} A deviation from the 2/3 behaviour has been observed in other systems as well.\cite{mgb2,ba-based} It is the central aim of this work to find out the distance over which the flux moves before a finite resistance appears across the transition. It is therefore important to know the exact form of the activation energy as we see next.

\begin{figure*}
\centering
\subfigure[]{
\includegraphics[width=7cm,height=5.5cm]{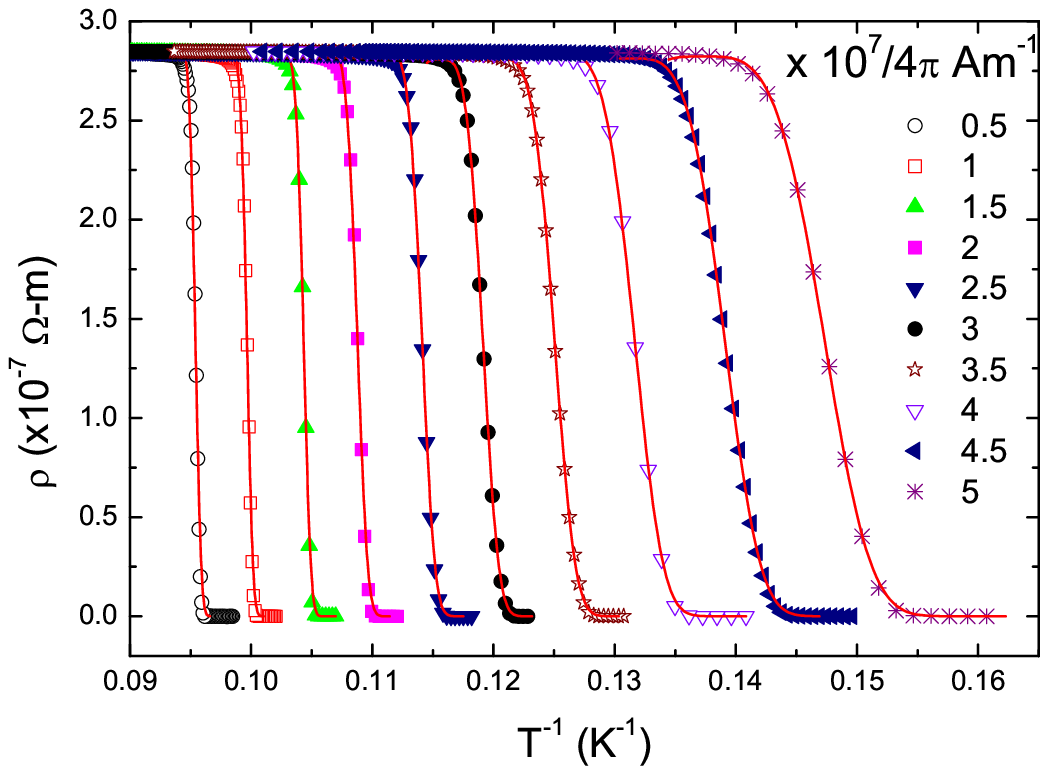}
\label{rt-non-arr}}
\hspace{0.5cm}
\subfigure[]{
\includegraphics[width=7cm]{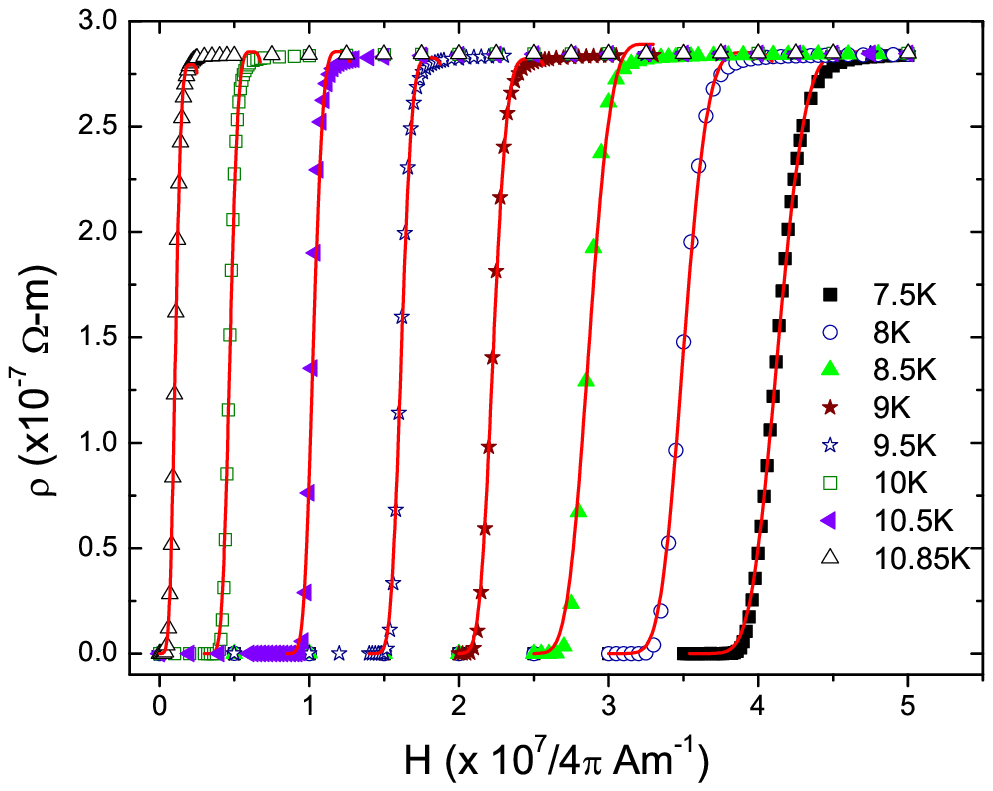}
\label{rh-non-arr}}
\caption{\label{non-arr}(a) Resitivity as a function of inverse temperature at various applied fields. The solid lines are fit to the non-Arrhenius relation mentioned in equation \ref{non-arr-rt-eqn}. (b) Resistivity as a function of field at various temperatures. The solid lines are fit to the non-Arrhenius relation mentioned in equation \ref{non-arr-rh-eqn}.}
\end{figure*}

Figure \ref{rt-non-arr} shows resistivity as a function of inverse temperature. The resistivity cannot be explained by a simple arrhenius relation where the activation energy is independent of temperature. Rather, it takes a non-arrhenius form which can be expressed as,\cite{kim-non-arr}
\begin{equation}
\rho(H,T) = \rho_0 e^{-U(H,T)/k_BT}
\label{rho}
\end{equation}
where
\begin{equation}
U(H,T) = U_0(H) (1-T/T_c)^\alpha
\label{U-H}
\end{equation}
Substituting equation \ref{U-H} in equation \ref{rho}, we get the temperature dependent resistivity at a particular field as, 
\begin{equation}
\rho(T) = \rho_0 \exp [-U_{(T=0)}(H)(1-T/T_c)^\alpha/k_BT]
\label{non-arr-rt-eqn}
\end{equation}

The best fit is obtained for $\alpha$ = 4 with a regression better than 0.996. Figure \ref{rh-non-arr} shows the resistive transition as a function of magnetic field at various temperatures. The magnetic field dependent isothermal resistivity can be expressed with a similar non-arrhenius form as,
\begin{equation}
\rho(H) = \rho_0 \exp [-U_{(H=0)}(T)(1-H/H_{C2})^\beta/k_BT]
\label{non-arr-rh-eqn}
\end{equation}
with the exponent $\beta$ once again = 4. The activation energies $U(H,T)$ determined from both temperature dependent resistivity in constant magnetic field and isothermal field dependent resistivity match quite closely as can be seen from table \ref{tab:activation}. The prefactor $U_0$ is quite large and is not the actual activation energy. Rather, it is the tangential slope of the arrhenius plot and is thus termed as the apparent activation energy.\cite{shove}

\begin{table*}[b]
	\caption{Activation energy determined from temperature dependent resistivity (R-T) in various fields and field dependent resistivity (R-H) at various temperatures. }
	\centering
		\begin{tabular}{c@{\hskip 1cm}c@{\hskip 1cm}c@{\hskip 1cm}c}
		\hline
			$T$ (K)&$\mu_0H$(T)&$U_{R-T}$(meV)&$U_{R-H}$(meV)\\\hline
			10&1&1.465&1.465\\ 
			9.5&1.5&3.374&3.671\\
			9&2&8.172&7.211\\
			8.5&3&0.0626&0.0626\\
			8&3.5&0.468&0.4775\\
			7.5&4&1.128&1.117\\
		\hline
		\end{tabular}
	
	\label{tab:activation}
\end{table*}

\begin{figure*}
\centering
\subfigure[]{
\includegraphics[width=7cm]{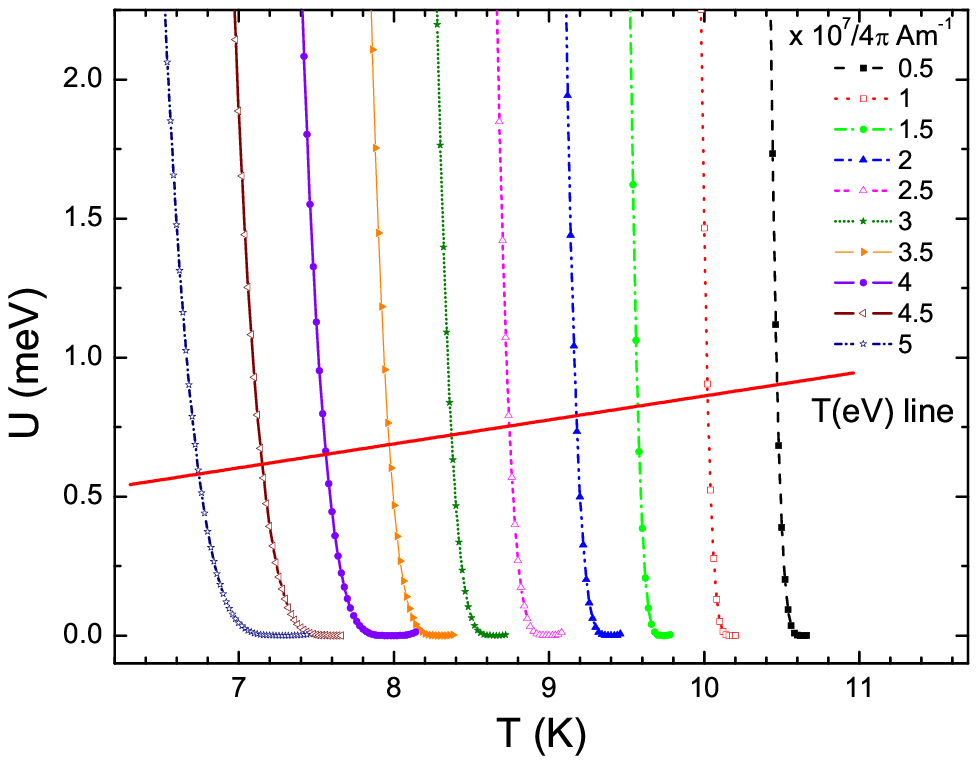}
\label{u-v-t}}
\hspace{0.5cm}
\subfigure[]{
\includegraphics[width=7cm,height=5.5cm]{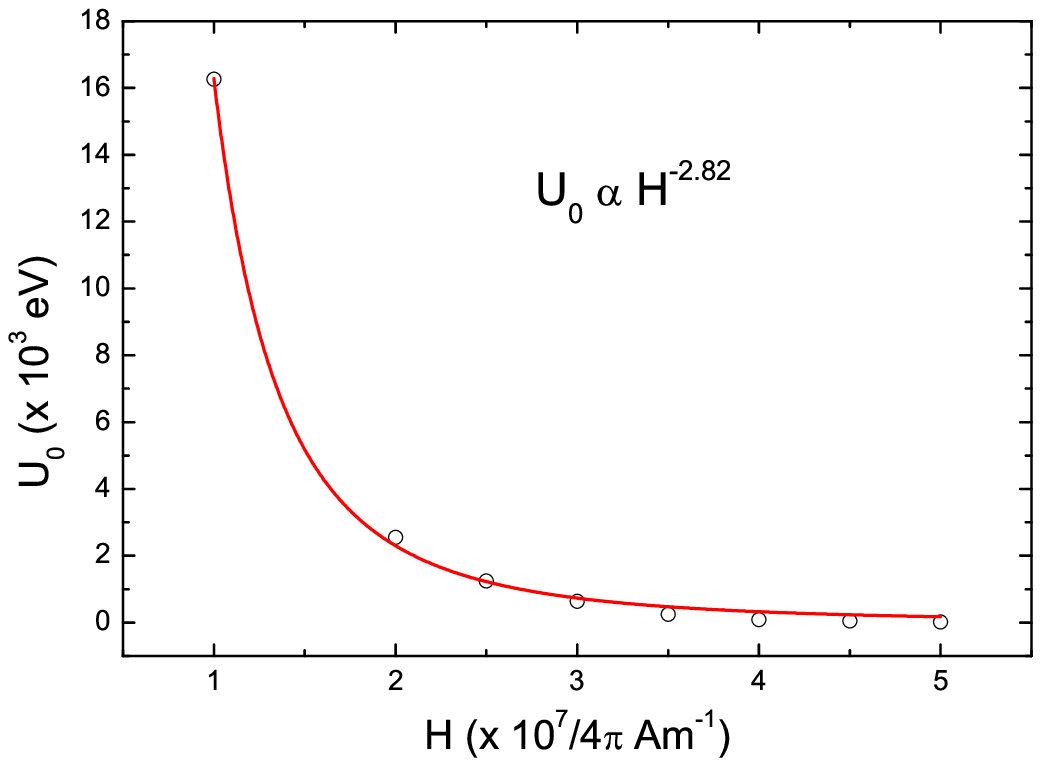}
\label{u0-v-h}}
\caption{\label{activation}(a) Activation energy as a function of temperature at various fields as determined from the fit of equation \ref{non-arr-rt-eqn} to the experimental data. (b) The prefactor $U_0$ in equation \ref{non-arr-rt-eqn} as a function of field.}
\end{figure*}

Figure \ref{u-v-t} shows the temperature dependent activation energy at various fields. The $T$(eV) line separates the flux creep and the thermally activated flux flow regions. When $U(T) << k_BT$, the thermally activated flux flow is the dominant phenomenon responsible for finite resistance. The variation of the prefactor $U_0(H)$ in equation \ref{U-H} is shown in figure \ref{u0-v-h}. The field dependence of $U_0$ can be explained as a power law with a very large exponent. We obtain a value of -2.82 for the exponent which is considerably larger than what has been seen in other systems.\cite{kim-non-arr} The significance of such a large value is not clear at present.

The non-Arrhenius shape of the resistive transition is similar to one of the universal characteristics exhibited by glass-forming liquids, which is the non-Arrhenius viscosity across the glass transition.\cite{shoving,shove} The Arrhenius temperature dependence of viscosity arises when there is a temperature independent barrier which has to be overcome (known as the activation energy) for the flow of molecules to occur. However there are only a few exceptions like SiO$_2$ and GeO$_2$ which exhibit an Arrhenius viscosity and in majority of cases, the activation energy increases upon cooling.\cite{shove} A non-Arrhenius viscosity can also arise in case of a ``landscape" of potential energy in the configuration space.\cite{landscape}

The \emph{shoving} model\cite{shoving} describes the non-Arrhenius temperature dependent viscosity of the glass forming liquids. We first briefly describe this model and then show how this model, which was originally proposed to explain the glass transition in molecular liquids, can be actually applied to the case of vortex matter to explain our observed results. We reproduce here some of the arguments by Dyre et. al\cite{shove} for the sake of continuity.     

The starting point of the shoving model is to view viscous liquids as `solids which flow'. Such a solid flows under force by \emph{sudden}, \emph{rare}, and \emph{localized} molecular rearrangements and extra volume is needed for the flow event to occur. The work done in creating this
extra volume is the activation energy and this is how the name of the model arose: in order to rearrange, the molecules must shove aside the surrounding molecules.\cite{shove} The elastic constant determining the shoving work is the shear modulus. The measure of how much the activation energy changes with temperature is the `index', $I$ = -dln$U$/dln$T$ and ranges between 2 and 6 for most of the glass forming liquids. In our case, the value of the index is 4 if we take the logarithmic derivative of the activation energy with respect to the reduced temperature as can be seen from equation \ref{U-H}.

The temperature dependent activation energy in the shoving model is given by 

\begin{equation}
U(T) = V_{corr} G(T)
\label{shove-activation}
\end{equation} 
where $V_{corr}$ is a \emph{correlation} volume and $G$ is the shear modulus.

The basic assumptions behind Eq. \ref{shove-activation} are as follows,\cite{shove}
\begin{enumerate}
\item The activation energy is (mainly) \emph{elastic} energy.
\item This elastic energy is located in (mainly) the \emph{surroundings} of the rearranging molecules.
\item The elastic energy is (mainly) \emph{shear} elastic energy, i.e., not associated with any density change.
\end{enumerate}

The model which was proposed for the molecular flow in viscous liquids (disordered solids) can be applied to the vortex matter if we replace the molecules with flux-lines. The flux-lines form a solid (lattice) under repulsion and thus have to shove aside the surrounding flux-lines if a flow has to take place. It is thus easy to see that the elastic energy is mostly located in the surroundings of the rearranging flux-lines. The elastic energy is also mainly the shear elastic energy as no density change is involved during the flow process because the density of flux lines is solely dependent on the applied field. Moreover, it costs a lot more energy to tilt the fluxons as we shall see later when we calculate the elastic constants of the flux-line lattice. Therefore the only drastic assumption needed to apply the shoving model to the flow of flux-lines is that the activation energy is mainly elastic energy. The Kramer model which explains the scaling laws of flux pinning implies this assumption while relating the strength of pinning with the appropriate elastic constant.\cite{kramer} In this work we build upon those assumptions and relate them with the molecular flow of disordered solids as described by the shoving model. We view the irreversible to reversible transition (or the resistive transition) in vortex matter as an event which can be described as the viscous flow of disordered solid. In that respect, the various pictures of glass-like behaviour,\cite{muller,fisher} flux creep,\cite{yeshurun,tinkham} or the melting of flux line lattice\cite{paradigm,melting} can be treated as equivalent, except for the shape of the irreversibility line, with the shear modulus of the vortex lattice as the key ingredient behind all these observed phenomenon. The apparent similarities (and differences) between the various interpretations of the resistive transition in vortex matter have been discussed in details by Brandt.\cite{similar}

For the following part of this article, the focus will be on equation \ref{shove-activation}. The activation energy is already obtained from resistivity measurements. Our task is now to obtain the shear modulus of vortex matter for our sample and identify a volume which can be used as the correlation volume. We use the magnetization measurements for this purpose.

\subsection{Magnetization measurements and determination of elastic constants of vortex matter}

\begin{figure}
\centering
\includegraphics[width=8cm]{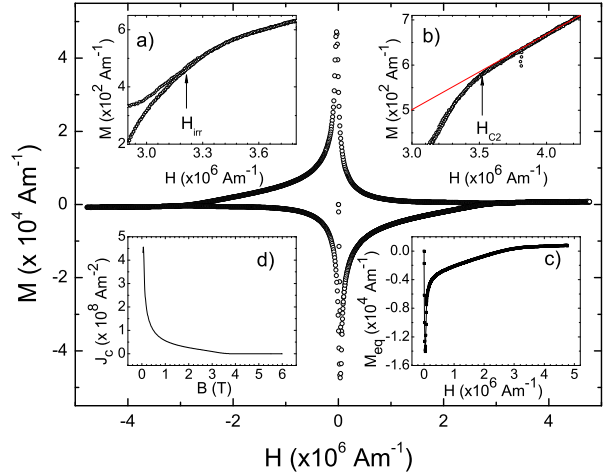}
\caption{\label{mh-7}Isothermal magnetization curve at 7K. The insets show the determination of various physical parameters from the magnetization measurement. (a) Determination of irreversibility field. (b) Determination of upper critical field $H_{C2}$. (c) The equilibrium magnetization as a function of field as determined from equation \ref{meq}. (d) The critical current density as a function of field as determined from equation \ref{jc}. } 
\end{figure}

Figure \ref{mh-7} shows the isothermal magnetization as a function of field of the Nb$_{75}$Zr$_{25}$ sample at T=7K. The insets show the various superconducting parameters obtained from this curve. The mixed state is paramagnetic at high fields and substantial positive magnetization can be observed even in the superconducting state. The upper critical field $H_{C2}$ is determined as the field value at which the magnetization deviates from the linear M-H curve, passing through origin, of the paramagnetic phase in the normal state. The irreversibility field is estimated from the opening of hysteresis between the field-increasing and field-decreasing curves.

The equilibrium magnetization is estimated as, 
\begin{equation}
M_{eq}(H) = \frac{M\uparrow(H)+ M\downarrow(H)}{2}
\label{meq}
\end{equation}

where $M\uparrow$ is the magnetization of the field-increasing cycle and $M\downarrow$ is the magnetization on the field-decreasing cycle.

The critical current density for the rectangular sample (in SI units) is estimated as,\cite{chen}
\begin{equation}
J_C(H) = 2 \times \frac{\Delta M(H)}{a_2(1-a_2/3a_1)}
\label{jc}
\end{equation}

where $\Delta M$ is the difference in magnetization between the field-decreasing and field-increasing curves at a particular field value and a$_1$ and a$_2$ are sample dimensions perpendicular to the direction of applied field, with $a_1$ $>$ $a_2$.

The thermodynamic critical field $H_C$ is calculated from the equilibrium magnetization as 

\begin{equation}
H_C^2 = 2\int_0^{H_{C2}} \! M_{eq}(H) \, \mathrm{d}H. 
\end{equation}

\begin{figure}
\centering
\includegraphics[width=8cm]{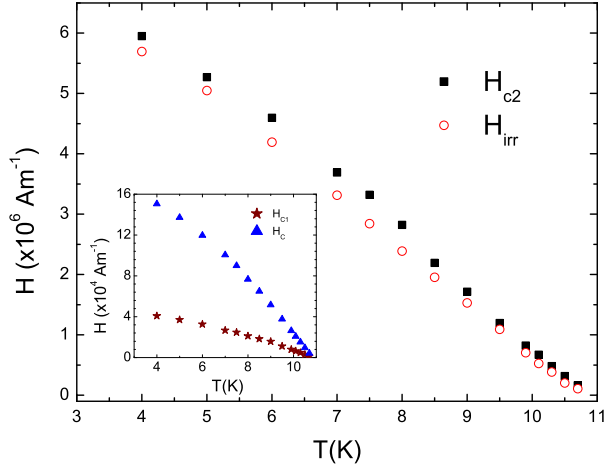}
\caption{\label{phase-dia}The field-temperature phase diagram of Nb$_{75}$Zr$_{25}$ which shows the upper critical field and the irreversibility line. Inset shows the lower critical field and the thermodynamic critical field as a function of temperature.} 
\end{figure}

The critical fields and the irreversibility line are plotted in figure \ref{phase-dia}. It should be noted that the irreversibility line does not follow the 2/3 dependence which is seen in case of the vortex glass\cite{muller} or the giant flux creep.\cite{yeshurun} Rather it is seen that $(1-T/T_C)$ $\propto$ $H_{irr}^{1.08}$, which does not match with any of the models to explain the resistive transition in vortex matter. However, as we have pointed out earlier, the shape of the irreversibility curve is not the focus of our present work and we are mainly concerned about the distance over which the flux lines move when a finite resistance appears across the transition.

Apart from the critical fields, the other useful quantities are the fundamental length scales, the coherence length $\xi$ and the penetration depth $\lambda$, which will be useful to understand the pinning properties when compared with the microstructure discussed earlier. 
The coherence length is estimated from the upper critical field as,
\begin{equation}
\xi^2 = \frac{\phi_0}{2\pi \mu_0 H_{C2}}
\end{equation} 

\begin{figure*}
\centering
\subfigure[]{
\includegraphics[width=7cm,height=5.5cm]{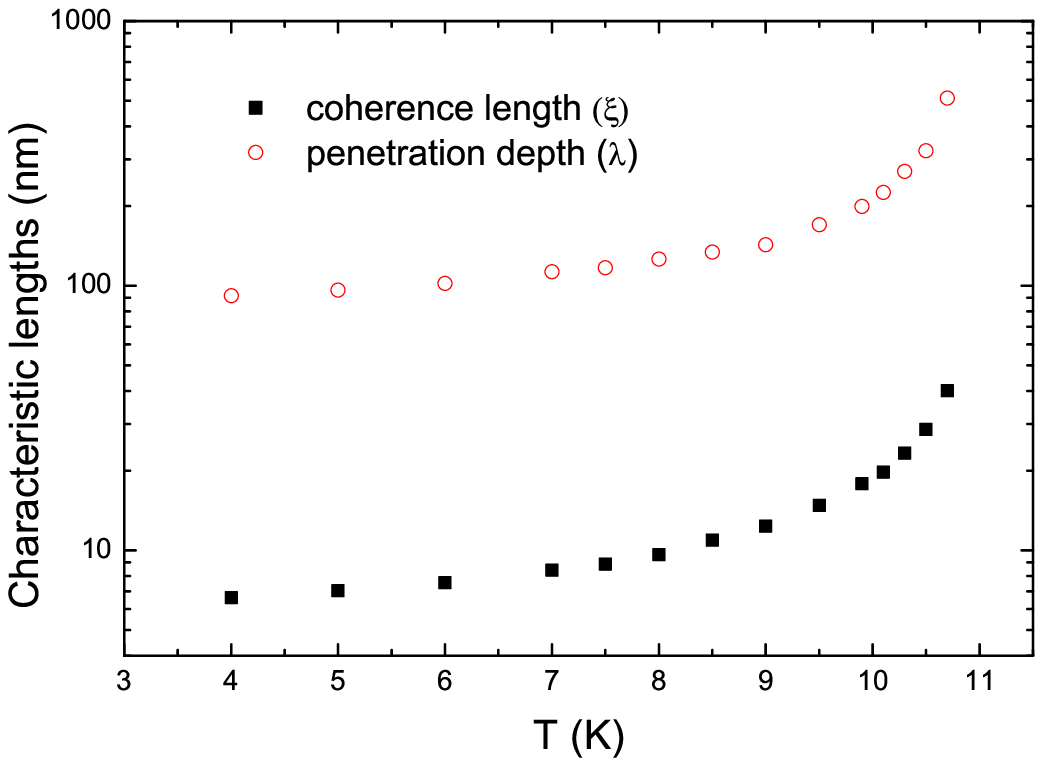}
\label{length}}
\hspace{0.5cm}
\subfigure[]{
\includegraphics[width=7cm]{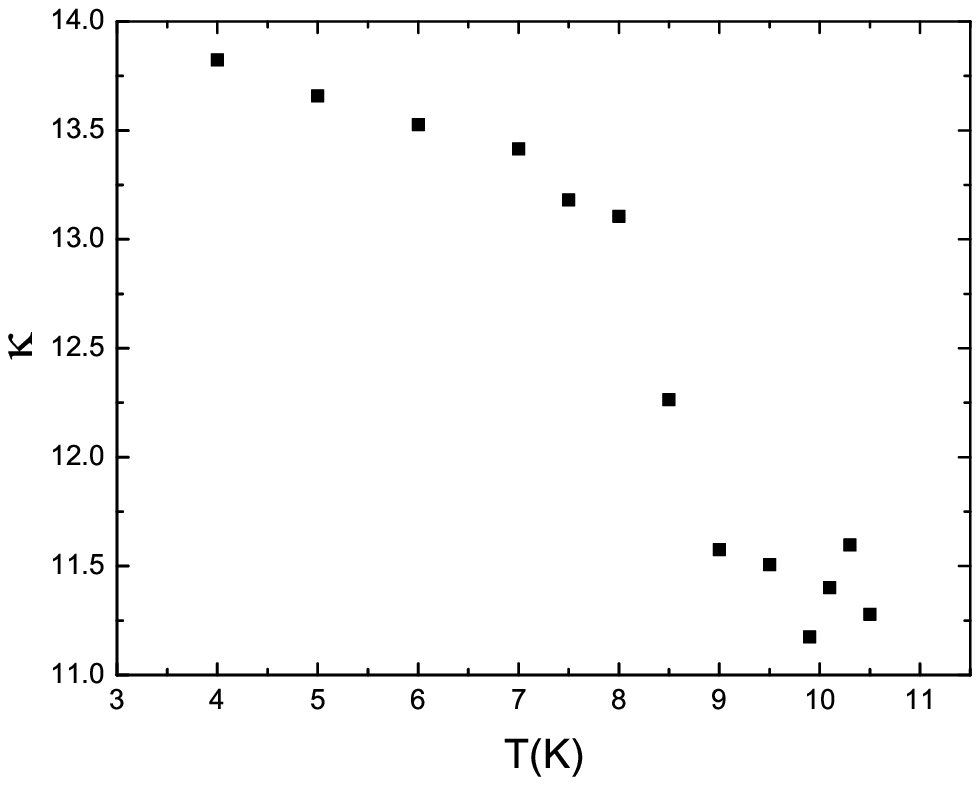}
\label{GL}}
\caption{\label{characteristic}(a) The characteristic length scales, penetration depth ($\lambda$) and coherence length ($\xi$) as a function of temperature and (b) The GL parameter ($\kappa$) as function of temperature for Nb$_{75}$Zr$_{25}$.}
\end{figure*}

The penetration depth is estimated from its relation to the lower critical field as,

\begin{equation}
H_{C1} = \frac{\phi_0 \mathrm{ln}(\kappa)}{4\pi \mu_0 \lambda^2}
\end{equation} 

where $\kappa$ is the GL parameter given as $\lambda$/$\xi$. The penetration depth, coherence length and the GL parameter are shown in figure \ref{characteristic}. 

The interesting point to be noted here is that the coherence length is quite large as compared to the typical defect sizes shown in figure \ref{tem}.  The penetration depth also being quite large, allows the field to penetrate over a large distance around the vortex core and the field may actually span a lot of defects before decaying.  It is therefore more likely to have more than one pinning mechanism as we see later during the description of the pinning properties. The large coherence length could also be the reason behind the sharp transition observed in bulk measurements. The coherence length for the composition which has the largest $T_C$ could turn the surrounding `normal' regions into superconducting regions through proximity effect, thereby making the composition inhomogeneity quite ineffective in broadening the transition. Figure \ref{GL} shows the variation of the GL parameter as a function of temperature. There is a sharp decrease in the value of $\kappa$ near 8K whose reason is not clear at present.

The disorder profile in our sample also gives rise to the possibility of collective pinning of the flux-lines. If the flux-lines are rigid and placed in a perfectly periodic arrangement, the pinning would be quite inefficient due to the random nature of the underlying pinning centres, which are in the form of composition variation, dislocations and amorphous regions of the dendritic arms (see figures \ref{sem-eds} and \ref{tem}). It is therefore favourable for the individual flux-lines to lower their energy by passing through the nearest pinning sites and deviate from the ideal periodic arrangement. This distortion (deformation) of the flux-lines however increases the elastic energy and the equilibrium flux-line configuration will be that arrangement which minimizes the sum of these two energies. This is the central theme of the theory of collective pinning by Larkin and Ovchinnikov.\cite{larkin} The distortion in the flux-line lattice can be then understood in terms of a certain correlation radius and correlation length within which the lattice remains reasonably undistorted. The elastic energy consists mainly of the shear and tilt experienced by the flux line lattice due to this distortion. The shear modulus $C_{66}$ is given by,\cite{brandt-elastic}
\begin{equation}
C_{66} = \frac{B_{C2}^2}{\mu_0}\frac{b(1-b)^2}{8\kappa^2}\left(1-\frac{1}{2\kappa^2}\right)(1-0.58b+0.29b^2)
\label{c66-eqn}
\end{equation}
where the reduced field $b$ = $B/B_{C2}$.
The tilt modulus $C_{44}$ is given by, 
\begin{equation}
C_{44} = BH_a
\end{equation}

\begin{figure*}
\centering
\subfigure[]{
\includegraphics[width=7cm]{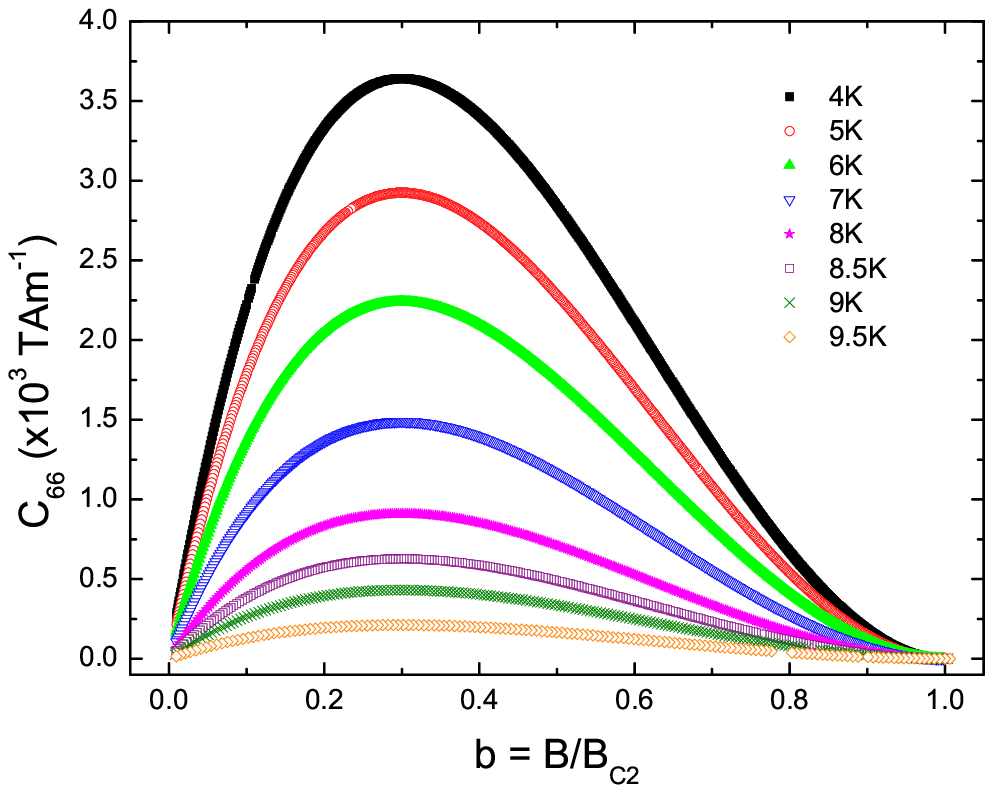}
\label{c66}}
\hspace{0.5cm}
\subfigure[]{
\includegraphics[width=7cm]{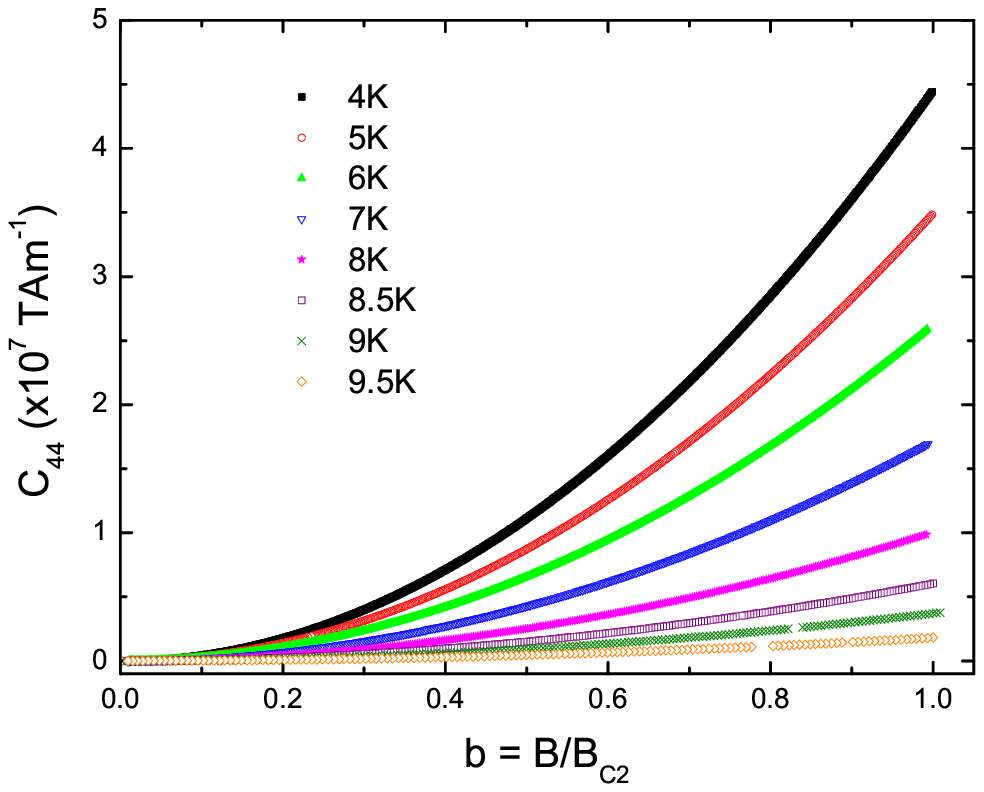}
\label{c44}}
\caption{\label{elastic}Elastic constants of vortex matter as a function of reduced field at various temperatures. (a) Shear modulus and (b) tilt modulus.}
\end{figure*}

Figure \ref{elastic} shows the shear modulus and the tilt modulus as a function of reduced field at various temperatures. It can be seen that the tilt modulus is nearly 3 to 4 orders of magnitude larger than the shear modulus for most of the field range. This implies that it is much easier to shear the flux lattice than to tilt it. This observation is important because the shoving model assumes that the flow occurs mostly by shearing the solid, which is the basis of equation \ref{shove-activation}.  

We shall first check the applicability of the collective pinning theory to the present Nb$_{75}$Zr$_{25}$ sample by studying the pinning properties. Once the applicability is justified, the correlation lengths obtained from the collective pinning theory along with the elastic constants of the flux line lattice will be used to solve equation \ref{shove-activation} in terms of the experimentally observed parameters.

\subsection{Pinning properties and the applicability of collective pinning theory}

To check if the pinning properties can be explained within the framework of single vortex pinning or collective pinning, the estimate of the depairing current is quite useful. The depairing current is given by,\cite{depairing} 

\begin{equation}
J_0 \sim \frac{4 B_C}{\sqrt{6}\mu_0 \lambda} = \frac{4 H_C}{\sqrt{6} \lambda}
\end{equation}

The value of the depairing current at 7K (estimated from figures \ref{phase-dia} and \ref{length}) turns out to be 4.98 x 10$^{11}$ Am$^{-2}$, with the highest value of J$_C$ being of the order of 10$^8$ Am$^{-2}$. 

The correlation length for single vortex pinning is,\cite{blatter-rmp}  

\begin{equation}
L_C^{sv} \sim \xi\left(\frac{J_0}{J_C}\right)^{1/2}
\end{equation}

which at 7K and 4T is 3.39 x 10$^{-4}$m.

The field region below which the single vortex pinning is applicable is,\cite{blatter-rmp} 

\begin{equation}
B_{sv} \sim \mu_0\left(\frac{J_C}{J_0}\right)H_{C2}
\end{equation}

which is 2.9 x 10$^{-7}$T at 7K. This field value is much smaller than the value of lower critical field. Moreover, the ratio of the critical (or the depinning) current density and the depairing current density is quite small, which allows us to discuss the pinning properties in terms of the weak collective pinning theory.\cite{blatter-rmp}  

\begin{figure*}
\centering
\subfigure[]{
\includegraphics[width=7cm]{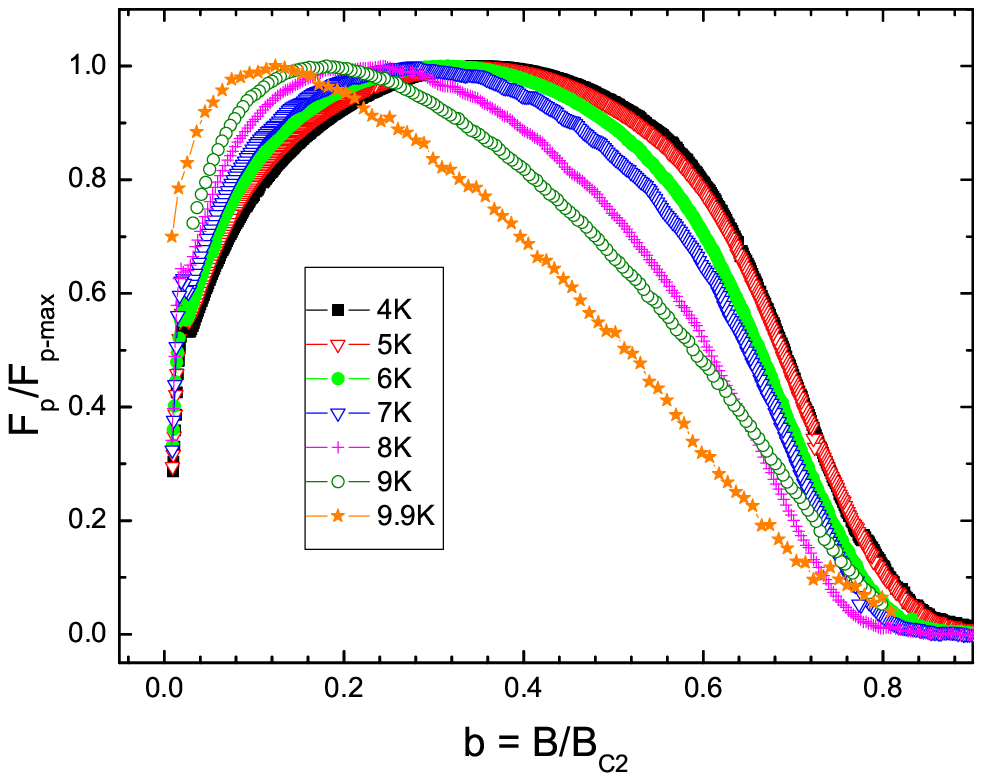}
\label{fp-norm}}
\hspace{0.5cm}
\subfigure[]{
\includegraphics[width=7cm]{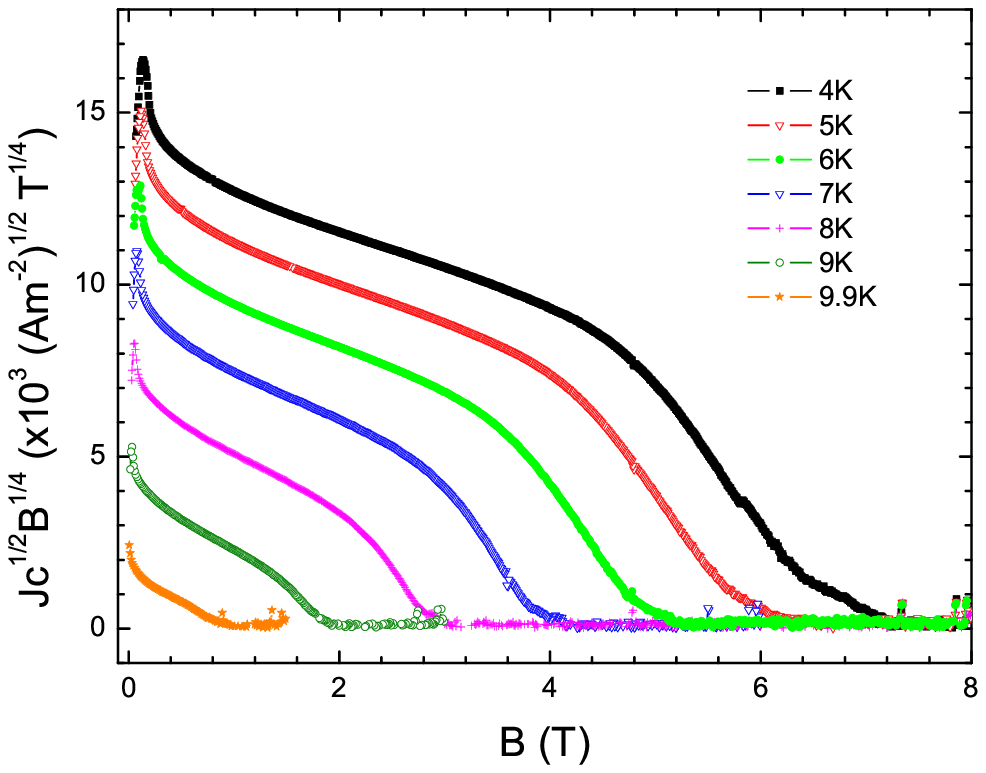}
\label{kramer}}
\caption{\label{norm-kramer}(a) Normalized pinning force density as a function of reduced field at various temperatures and (b) Kramer plot .} 
\end{figure*}

Figure \ref{fp-norm} shows the normalized pinning force density as a function of reduced field.  Two distinct temperature regimes can be identified in fig. \ref{fp-norm} where the shape of the pinning curves are quite different. At temperatures below nearly 8K, the curve is quite broad and appears to be a combination of two pinning laws. At temperatures above 8K, the peak in the curve shifts to much lower field values. The distinction of the temperature regimes also coincide with the sharp change in the GL parameter as a function of temperature (see fig. \ref{GL}). Similar shift in the peak of the pinning curves with temperature has been observed in case of polycrystalline MgCNi$_3$ with graphite nanoprecipitates.\cite{nano-precip} The Kramer plot is shown in figure \ref{kramer}. The kramer plot ($J_C^{1/2}B^{1/4}$ as a function of $B$) is piecewise linear with two major straight line portions which indicates that there are at least two dominant pinning mechanisms present in our sample. The normalized pinning force density can be scaled with the combination of two pinning laws as,

\begin{equation}
f(p) = \frac{F_p}{F_{p_{max}}}= C_1b^{p_1}(1-b)^{q_1} + C_2b^{p_2}(1-b)^{q_2}
\label{2-pin}
\end{equation}

\begin{figure*}
\centering
\subfigure[]{
\includegraphics[width=7cm]{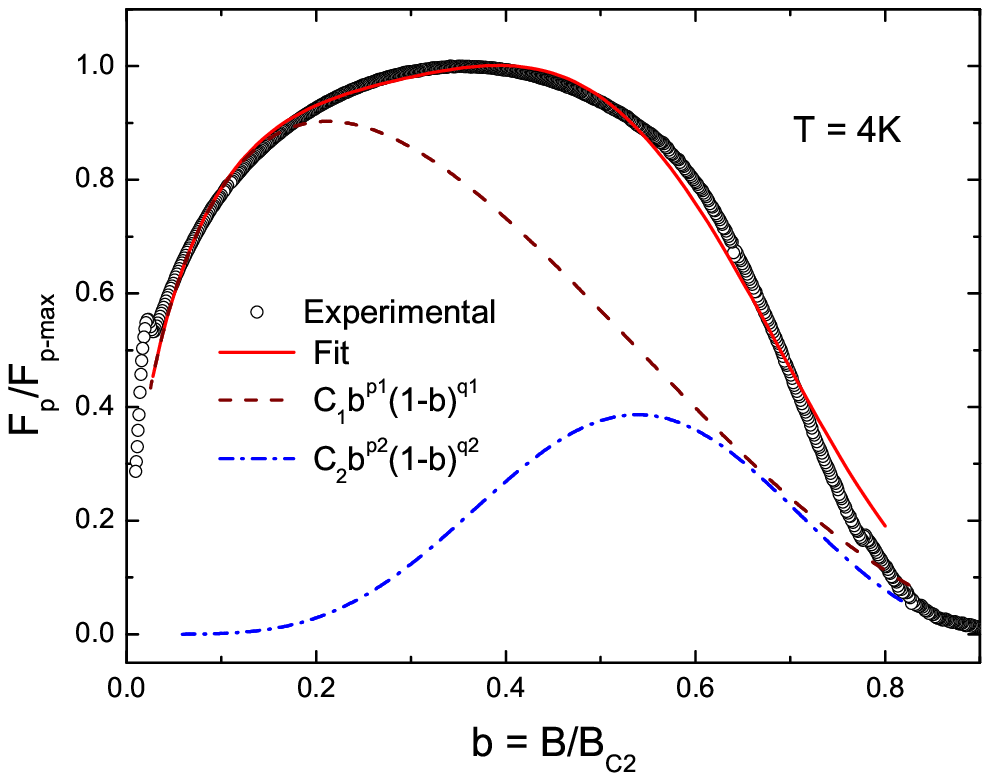}
\label{fp-norm-4k}}
\hspace{0.5cm}
\subfigure[]{
\includegraphics[width=7cm]{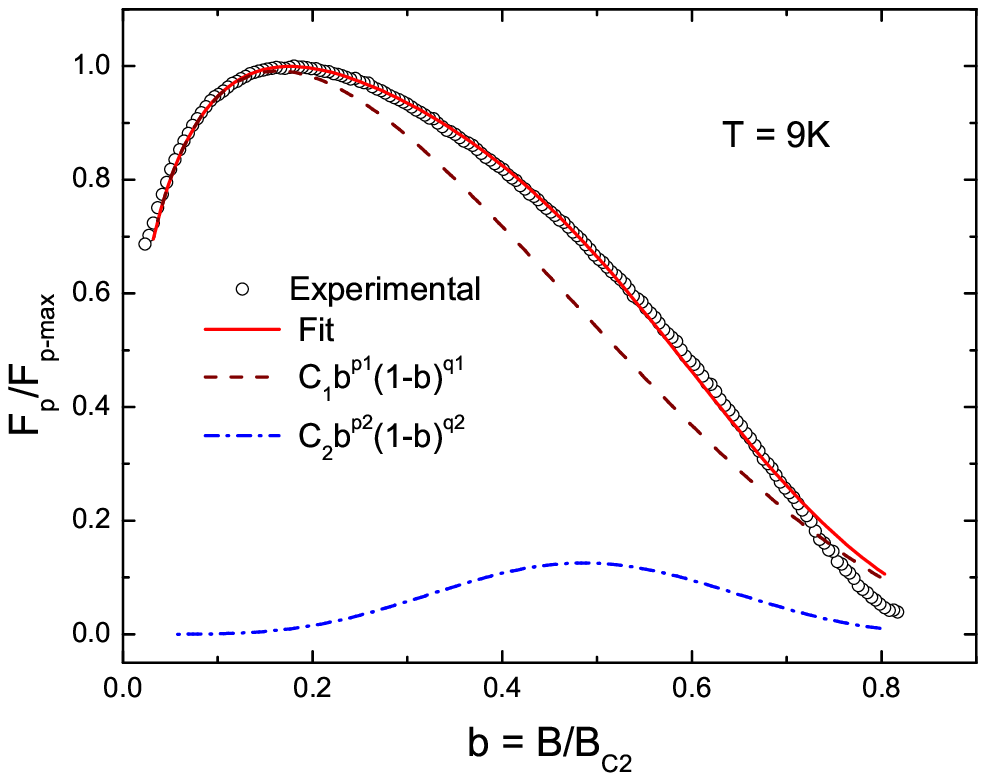}
\label{fp-norm-9k}}
\caption{\label{fp-fit}The normalized pinning force as a function of reduced field at two different temperatures along with the fit to equation \ref{2-pin}. The total of the two scaling laws and the individual components are plotted along with the experimental data.}
\end{figure*}

Figure \ref{fp-fit} shows the normalized pinning force density at two temperatures, one far below $T_C$ and one at a higher temperature near $T_C$. The pinning curves can be fitted quite well with equation \ref{2-pin}. Such a combination has been used earlier to explain the pinning curve in Chevrel phase superconductors.\cite{bonney} It was shown that two pinning mechanisms were possible within the framework of collective pinning depending upon the relationship between the grain size and the correlation radius.\cite{bonney}  The correlation radius in the collective pinning theory is given by,\cite{larkin}
\begin{equation}
R_c = \sqrt{\frac{2a_{fll}C_{66}}{BJ_C}}
\end{equation}
and the correlation length is
\begin{equation}
L_c = 2 \times \sqrt{\frac{a_{fll}C_{44}}{BJ_C}}
\end{equation}
where,
\begin{equation}
a_{fll} = 1.075 \times \sqrt{\frac{\phi_0}{B}}
\end{equation}
is the flux line lattice parameter for a triangular Abrikosov lattice.\cite{tinkham-book}

\begin{figure*}
\centering
\subfigure[]{
\includegraphics[width=7cm]{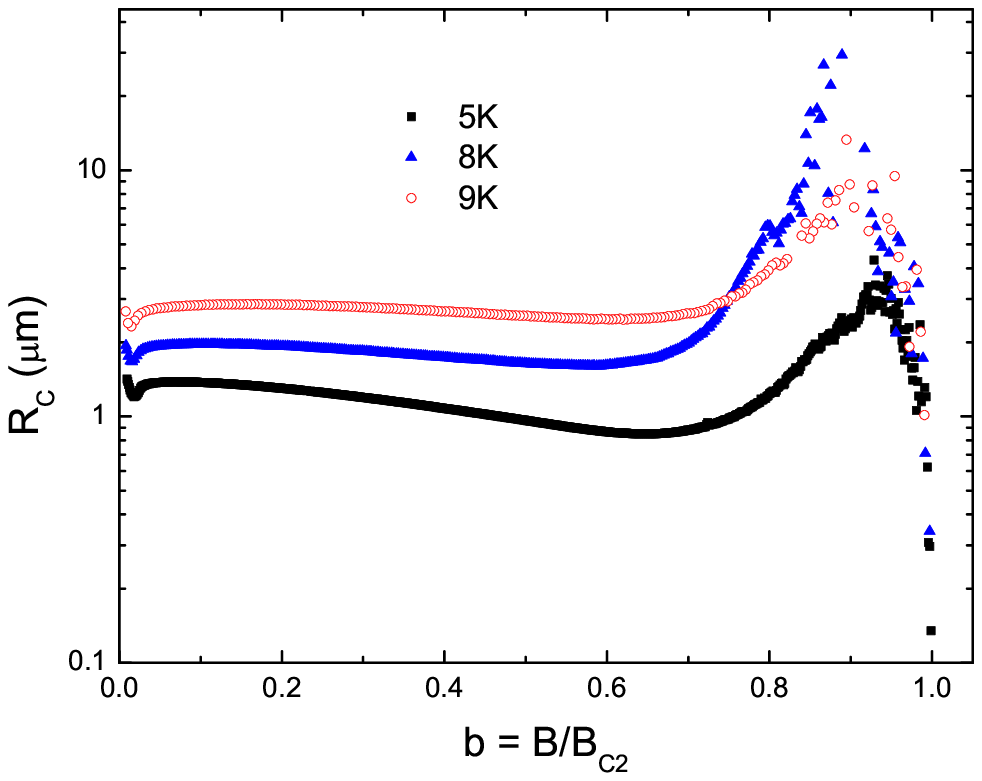}
\label{rc}}
\hspace{0.5cm}
\subfigure[]{
\includegraphics[width=7cm]{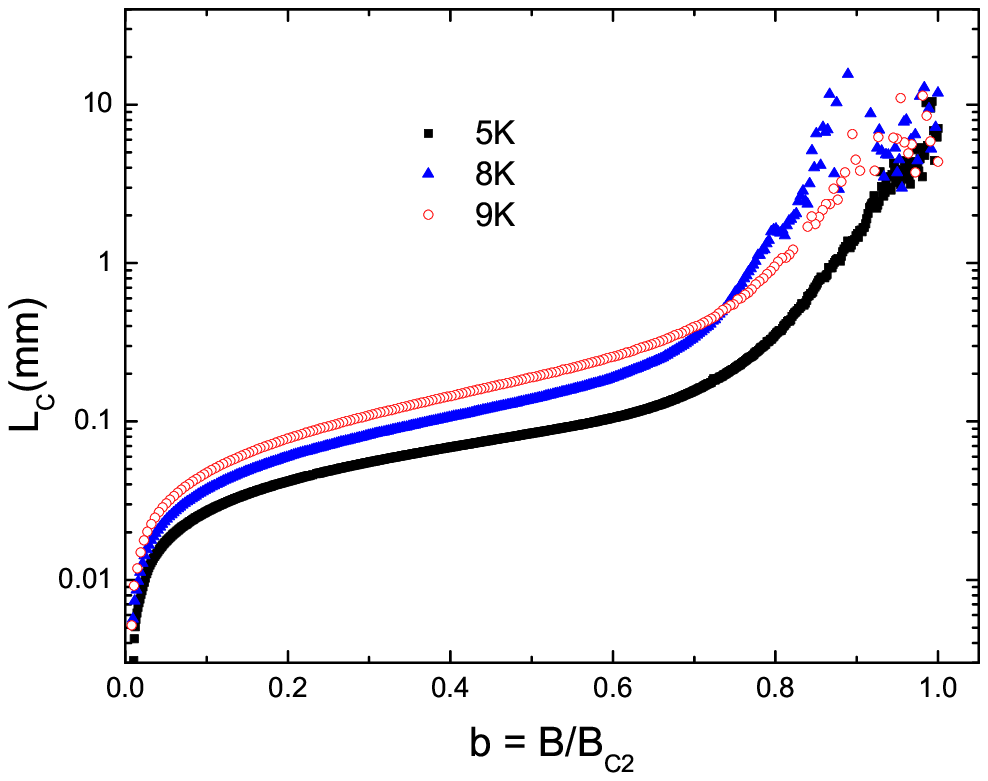}
\label{lc}}
\caption{\label{correlation}(a) The correlation radius and (b) the correlation length as a function of reduced field at three representative temperatures.}
\end{figure*}

The variation of the correlation radius $R_C$ and the correlation length $L_C$ as a function of reduced field is shown in figure \ref{correlation} at three representative temperatures. The correlation radius is in a plane perpendicular to the direction of applied field and the correlation length is along the flux line. As we have seen earlier (fig. \ref{elastic}), the tilt modulus is much larger than the shear modulus and hence the pinning properties could be quite insensitive to the correlation length and mainly depend on the relative changes of the correlation radius.  The correlation radius remains fairly constant till a reduced field of about 0.7 at all temperatures. At slightly higher fields, there is a drastic increase by almost an order of magnitude before $R_C$ tends to drop down to zero at $B_{C2}$. The observed two pinning regimes can be explained with this rapid change in $R_C$ as we discuss next.

Figure \ref{fp-fit} shows the normalized pinning force density at 4K and 9K each fitted with equation \ref{2-pin}. The normalized pinning force at 4K (fig. \ref{fp-norm-4k}) clearly shows the presence of two pinning mechanisms, one with the $p_1$ $\approx$ 0.5 and $q_1$ $\approx$ 2, which is commonly observed for grain boundary pinning or surface pinning\cite{dew-hughes} with a peak at b $\approx$ 0.2. The second component of equation \ref{2-pin} gives quite high values of $p_2$ and $q_2$. The fitting is done till $b$ $\approx$ 0.8 by following the conventionally accepted procedure as the inhomogeneities in the sample significantly affect the tails of the pinning curves.\cite{godeke} Such high values of the exponents have been reported in other systems as well and are thought to arise due to inhomogeneity in the sample.\cite{bonney} Treating the exponents $p_2$ and $q_2$ as free running parameters in the fitting procedure yields the best fit as seen in case of Chevrel phase compounds\cite{bonney} and Ti-V alloys\cite{matin} for example.      

At higher temperatures (fig. \ref{fp-norm-9k}) the peak of the pinning curve distinctly shifts towards b $\approx$ 0.2 and the surface (or the grain boundary) pinning appears to be the dominant mechanism of flux pinning. 

The answer to why the surface pinning becomes more dominant at higher temperatures can be obtained from the behaviour of $R_C$ as a function of temperature and its relation with the microstructure of the sample. From figure \ref{rc} it can be seen that $R_C$ changes by almost an order of magnitude (especially at higher fields) as a function of temperature. At lower temperatures ($\approx$ 5K and below), $R_C$ is nearly 1 $\mu$m.  With this correlation radius, the defects and composition inhomogeneity within the dendritic arms (see fig. \ref{sem-eds}) can offer effective pinning centres like the intragranular defects in case of Chevrel phase superconductors.\cite{bonney} The dendritic arms which have a higher Zr concentration offer additional surfaces similar to grain boundaries in the average sample matrix. These additional surfaces could be the cause behind the usual $p_1$ = 0.5 and $q_1$ = 2 pinning law even when there are no well defined grains in our sample as we have seen during the discussion on microstructure earlier. The inhomogeneities within the dendritic arms could be the probable reason for the higher values of exponents $p_2$ and $q_2$ in equation \ref{2-pin}. At higher temperatures (and higher fields), the correlation radius grows by nearly 8 to 10 times (see fig. \ref{rc}) and is probably much larger than the internal structure of the dendritic arms. The inhomogeneities and defects within the dendritic arms are thus averaged out which makes the pinning over smaller length scales quite ineffective. This explains the shift of the peak in the pinning curve to lower fields and the dominance of the surface pinning at higher temperatures.    

The values of the fitting parameters along with the correlation radii at the peak reduced-field values $b_{max}$ of the pinning curves are given in table \ref{tab:pinfit}. It is thus justified that the theory of collective pinning can be applied to our sample and we now proceed to use this framework in equation \ref{shove-activation} to answer the central question posed in the introduction.

\begin{table*}
	\caption{The fitting parameters in equation \ref{2-pin} along with the correlation radii at the peak reduced-field values of the pinning curves.}
\vskip 0.2cm	
	\centering
		\begin{tabular}{|c|@{\hskip 0.2cm}c@{\hskip 0.5cm}c@{\hskip 0.5cm}c@{\hskip 0.5cm}c@{\hskip 0.5cm}c|@{\hskip 0.2cm}c@{\hskip 0.5cm}c@{\hskip 0.5cm}c@{\hskip 0.5cm}c@{\hskip 0.5cm}c|}
		\hline
		&\multicolumn{5}{c|@{\hskip 0.2cm}}{Low field peak}&\multicolumn{5}{c|}{High field peak}\\	
		
		$T$(K)&$C_1$&$p_1$&$q_1$&$b_{max}$&$R_c$($\mu$m)&$C_2$&$p_2$&$q_2$&$b_{max}$&$R_c$($\mu$m)\\
		\hline
		4&3.453&0.55&2.05&0.21&1.21&230.5&4.99&2.27&0.539&0.85\\
		
		5 &3.281 &0.52 &1.95 &0.211 &1.29 &183.2 &4.95 &4.15 &0.543 &0.917\\
		6 &3.414 &0.52 &2.05 &0.206 &1.424 &231.4 &5 &4.55 &0.524 &1.064\\
		7 &3.268 &0.48 &2.05 &0.189 &1.59 &240.8 &4.99 &4.80 &0.509 &1.241\\
		7.5 &3.275 &0.48 &2 &0.193 &1.761 &284.5 &5.05 &5.50 &0.478 &1.462\\
		8 &3.397 &0.48 &2.05 &0.191 &1.932 &133.0 &4.99 &5.05 &0.496 &1.644\\
		8.5 &3.401 &0.48 &2.05 &0.188 &2.24 &141.6 &4.99 &5.06 &0.496 &1.93\\
		9 &2.949 &0.4 &2.05 &0.162 &2.856 &157.6 &5 &5.30 &0.483 &2.575\\
				
		\hline
		\end{tabular}
	
	\label{tab:pinfit}
\end{table*}

\subsection{Estimating the distance of flux movement during resistive transition}

We now compare the estimates of the displacement of the flux-line lattice by considering two processes of flux movement. One is the process of flux creep and the other is the process of the flow of viscous liquid as described in the shoving model.

The pinning properties have shown that the influence of quenched disorder can be explained in terms of collective pinning theory. To estimate the distance of flux movement when a finite resistance appears across the resistive transition, it is important to first know the volume of the flux bundle which is involved in the displacement. A collectively pinned object is classified as \emph{large-bundle} when $R_C$ $>$ $\lambda$, which seems to be the case in the present situation.\cite{blatter-rmp} 

The total elastic energy involved during the creep process can be estimated within the idea that a bundle of bundles, a superbundle with dimensions $R_\parallel$ (parallel to the jump direction), $R_\perp$ = $R_C$, (transverse to the field and the jump direction), and $L^b$ (along the field direction), will constitute the elementary unit.\cite{blatter-rmp} (Details can be seen in figure 19 of Blatter et al.\cite{blatter-rmp}) By taking into account the compression of the neighbouring volume of the flux bundle during the jump process, one obtains a scaling relation,\cite{blatter-rmp}
 
\begin{equation}
R_\parallel = \frac{\lambda}{a_{fll}}R_\perp
\end{equation}

When the tilt and compression energies are similar, $R_\parallel$ $\simeq$ $L^b$, which are the longest dimensions of the superbundle.\cite{blatter-rmp}

For example, at 8K and 2.9T in the case of the present sample, $R_\perp$ $\simeq$ $R_C$ = 6.042 $\mu$m and  $R_\parallel$ $\simeq$ $L^b$ = 26.59 $\mu$m. Similar order of magnitudes for these characteristic length scales have been actually experimentally estimated on another conventional low-T$_C$ material 2$H$-NbSe$_2$ by using neutron scattering.\cite{yaron}

The activation energy of a superbundle in the creep process near J$_C$ is given by\cite{blatter-rmp}

\begin{equation}
U_c^b \simeq C_{66}\frac{u^2}{R_\perp^2}R_\perp R_\parallel L^b
\end{equation}

where $u$ is the displacement, which we term as $\Delta x$ from here onwards. 

We wish to emphasize here that these are only order of magnitude estimates and the numerical factors have been dropped out from the calculations.\cite{blatter-rmp} For estimating the displacement of the flux bundle during the creep process, the activation energy estimated from resistivity measurements and the shear modulus estimated from the magnetization measurements is used. The typical length scales mentioned in these calculations and the associated displacement at various field and  temperature values, when a finite resistance appears just above the noise floor across the resistive transition are given in table \ref{tab:disp}. 

To solve equation \ref{shove-activation} we need the activation energy $U$, the correlation volume $V_{corr}$ and the shear modulus $C_{66}$. The $V_{corr}$ should not be confused with the correlation volume $V_C$ as mentioned in the theory of collective pinning.\cite{larkin}

The correlation volume during shoving is given by the shoving model as,\cite{shoving}

\begin{equation}
V_{corr} \simeq (\Delta V)^2/V
\label{v-corr}
\end{equation}

Where $V_{corr}$ = $U$/$C_{66}$ from the shoving model (equation \ref{shove-activation}). (we have dropped the 2/3 factor arising due to spherical symmetry to get only an order of magnitude estimate) $V$ is  volume indulged in shoving and $\Delta V$ is the volume change during shoving. We have taken $U$ to be same as $U_c^b$ which is obtained earlier from fig. \ref{non-arr} The task is now to identify the volume $V$ which is involved in shoving. The natural choice of such a volume within the collective pinning theory would be a parallelepiped of sides $R_C$ (or $R_\perp$) and length $L^b$, thereby giving a volume of $R_C^2$$L^b$ which would move when an external force is applied to it. In the shoving model, the shear energy is the only elastic energy related to the activation energy and thus the compression of the flux-line lattice is neglected. Therefore both the sides of the parallelepiped, $R_\perp$ and $R_\parallel$ are taken to be same. 
 
To know about the change in volume during the shoving process, (i.e. the distance over which flux lattice moves) we have chosen those temperature and field values after which a finite resistance appears just beyond the noise floor during resistivity measurements. As the value of $J_C$ enters into the calculation of $R_C$, the first non-zero value of $J_C$ beyond noise floor during magnetization measurements is used where, strictly speaking, the sample is not in the resistive state. Moreover, the appearance of finite resistance is also a function of the sensing current through the sample. Therefore, the calculation is only an order of magnitude estimate of the actual phenomenon and should be treated likewise.   

At $T$ = 9K and $\mu_0H$ = 1.9T, we have $R_C$ $\approx$ 7.541 $\mu$m and $L^b$ $\approx$ 29.82 $\mu$m. This gives,
\begin{equation}
V = R_C^2 L^b \approx 16.95 \times 10^{-16} \; m^3
\end{equation}

At these temperature and field values, $U$ = 2.47 $\times$ 10$^{-21}$J and $C_{66}$ = 30.55 TAm$^{-1}$ (or Jm$^{-3}$). This gives from equation \ref{shove-activation}, $V_{corr}$ = 8.08 $\times$ 10$^{-23}$ m$^3$.

From equation \ref{v-corr} above, 

\begin{equation}
(\Delta V)^2 \simeq V_{corr} \times V \approx 1.37\times 10^{-37}m^6 
\end{equation}

This gives $\Delta V$ $\approx$ 3.7$\times$10$^{-19}$ m$^3$. The change in area during the shoving process would therefore be $\Delta V$/$L^b$ which is nearly 124.13$\times$10$^{-16}$ m$^2$.
From simple geometric considerations, when the flux-line lattice shoves the neighbouring area aside by a distance of $\Delta x$, the change in area $\Delta V$/$L^b$ for a square of sides $R_C$ turns out to be, 

\begin{equation}
4 (\Delta x)^2 + 4 \Delta x R_C
\end{equation} 

This gives $\Delta x$ $\approx$ 4.11$\times$10$^{-10}$m. 

The displacement estimated from the superbundle creep picture turns out to be 8.27$\times$10$^{-10}$m which is nearly double but of the same order of magnitude as estimated by using the shoving model.

Table \ref{tab:disp} shows the comparative study of the flux-line lattice displacement within the shoving model\cite{shoving} (i.e. if the vortex matter is treated similar to the viscous flow of disordered solid) and the superbundle creep picture which is more conventionally used to explain the resistive transition in vortex matter\cite{blatter-rmp} at various fields and temperatures at which the irreversibility transition occurs. We see that in all cases, the estimates from both the pictures have the same order of magnitude. It should be noted that these are only order of magnitude estimates and the actual numerical factors are required (not available within theory, to the best of our knowledge) to compare these displacements with the flux-line lattice constant.

The implications of our study are quite interesting. We have attempted to find the analogies between the flow of disordered solids or viscous liquids with the resistive transition in vortex matter. We have not assumed any model like the superconducting glass or lattice melting for explaining the shape of the irreversibility line. Our arguments are based purely on thermodynamic considerations of elastic energies and could thus be a special case of a more general phenomenon of pinning and depinning properties elastic manifolds in random media.\cite{fisher-random} 

\begin{table*}
	\caption{The comparison between the displacement of flux line lattice estimated from the shoving model and within the superbundle creep picture at various temperature and field values along the irreversibility line.}
\vskip 0.2cm	
	\centering
		\begin{tabular}{c@{\hskip 0.3cm}c@{\hskip 0.3cm}c@{\hskip 0.3cm}c@{\hskip 0.3cm}c@{\hskip 0.3cm}c@{\hskip 0.3cm}c@{\hskip 0.3cm}c@{\hskip 0.3cm}c@{\hskip 0.3cm}c@{\hskip 0.3cm}c}
		\hline
			$T$ &$\mu_0H$&$U$&$C_{66}$&$R_\perp$&$R_\parallel$&$L^b$&$L_C$&a$_{fll}$&$\Delta x_{(shoving)}$&$\Delta x_{(bundle)}$\\
			(K) &(T)& (10$^{-21}$J)& (T.Am$^{-1}$) & ($\mu$m) &($\mu$m)&($\mu$m)&(mm)&(nm)&(nm)&(nm)\\
		\hline
		9.5&1.3&4.35&18.28&6.06&24.1&24.1&2.33&42.7&\textbf{0.78}&\textbf{1.57}\\
		9&1.9&2.47&30.55&7.54&29.8&29.8&3.27&35.4&\textbf{0.41}&\textbf{0.83}\\
		8.5&2.3&2.84&80.84&4.68&18.8&18.8&1.51&32.2&\textbf{0.34}&\textbf{0.68}\\
		8&2.9&2.54&143.64&6.04&26.6&26.6&1.84&28.7&\textbf{0.20}&\textbf{0.39}\\
		7.5&3.6&1.28&118.11&8.18&37.2&37.2&3.42&25.7&\textbf{0.13}&\textbf{0.25}\\
		7&4&3.10&138.77&7.35&34.0&34.0&3.15&24.4&\textbf{0.20}&\textbf{0.38}\\
		6&5&2.89&185&4.2&19.6&19.6&1.95&21.8&\textbf{0.22}&\textbf{0.41}\\
		\hline
		\end{tabular}
	
	\label{tab:disp}
\end{table*}

\section{Conclusion}

In conclusion, we have studied the resistive transition of vortex matter in highly strained sample of Nb$_{75}$Zr$_{25}$. The detailed study of microstructure of the sample enabled us to apply the theory of collective pinning to understand the pinning properties of flux line lattice in our sample and determine the elastic constants of vortex matter. The non-Arrhenius shape of the resistive transition, both as a function of temperature and magnetic field, showed that the models of viscous flow of disordered solids can be indeed applied to vortex matter. The arguments were based on purely elastic energy considerations instead of assuming any particular model or shape of the irreversibility line in the field-temperature phase space. Our results show that the viscous flow of disordered solids and the flux-creep phenomenon in hard type-II superconductors could be the manifestation of same underlying physical principles. These studies should provide sufficient interesting inputs for further experiments which can image the actual flow of flux line lattice and those theories which treat the flux flow in type-II superconductors as special case of the general phenomenon of plastic depinning of driven systems. 

\section*{Acknowledgments}

The authors wish to acknowledge Dr. L. S. Sharath Chandra for help with magnetization measurements, Dr. Tapas Ganguli for help with the x-ray diffraction measurements and Dr. A. K. Srivastava for guidance on the transmission electron microscopy measurements.

\end{document}